\pdfoutput=1
\documentclass[8.5pt,twoside,twocolumn]{article}
\usepackage[super,sort&compress,comma]{natbib} 
\usepackage{mhchem}
\usepackage{times,mathptmx}
\usepackage{sectsty}
\usepackage{balance} 
\usepackage{color} 

\usepackage{amsmath}
\usepackage{amssymb}
\usepackage{float}
\newcommand{\e}[1]{\times10^{#1}}
\newcommand{\gd}{\dot{\gamma}}

\usepackage{graphicx} %eps figures can be used instead
\usepackage{lastpage}
\usepackage[format=plain,justification=raggedright,singlelinecheck=false,font=small,labelfont=bf,labelsep=space]{caption} 
\usepackage{fancyhdr}

\newif\ifold
\begin{document}

\thispagestyle{plain}
\fancypagestyle{plain}{
%\fancyhead[L]{\includegraphics[height=8pt]{headers/LH}}
%\fancyhead[C]{\hspace{-1cm}\includegraphics[height=20pt]{headers/CH}}
%\fancyhead[R]{\includegraphics[height=10pt]{headers/RH}\vspace{-0.2cm}}
\renewcommand{\headrulewidth}{1pt}}
\renewcommand{\thefootnote}{\fnsymbol{footnote}}
\renewcommand\footnoterule{\vspace*{1pt}% 
\hrule width 3.4in height 0.4pt \vspace*{5pt}} 
\setcounter{secnumdepth}{5}

\makeatletter 
\def\subsubsection{\@startsection{subsubsection}{3}{10pt}{-1.25ex plus -1ex minus -.1ex}{0ex plus 0ex}{\normalsize\bf}} 
\def\paragraph{\@startsection{paragraph}{4}{10pt}{-1.25ex plus -1ex minus -.1ex}{0ex plus 0ex}{\normalsize\textit}} 
\renewcommand\@biblabel[1]{#1}            
\renewcommand\@makefntext[1]% 
{\noindent\makebox[0pt][r]{\@thefnmark\,}#1}
\makeatother 
\renewcommand{\figurename}{\small{Fig.}~}
\sectionfont{\large}
\subsectionfont{\normalsize} 

%\fancyfoot{}
%\fancyfoot[LO,RE]{\vspace{-7pt}\includegraphics[height=9pt]{headers/LF}}
%\fancyfoot[CO]{\vspace{-7.2pt}\hspace{12.2cm}\includegraphics{headers/RF}}
%\fancyfoot[CE]{\vspace{-7.5pt}\hspace{-13.5cm}\includegraphics{headers/RF}}
%\fancyfoot[RO]{\footnotesize{\sffamily{1--\pageref{LastPage} ~\textbar  \hspace{2pt}\thepage}}}
%\fancyfoot[LE]{\footnotesize{\sffamily{\thepage~\textbar\hspace{3.45cm} 1--\pageref{LastPage}}}}
\fancyhead{}
\renewcommand{\headrulewidth}{1pt} 
\renewcommand{\footrulewidth}{1pt}
\setlength{\arrayrulewidth}{1pt}
\setlength{\columnsep}{6.5mm}
\setlength\bibsep{1pt}

\twocolumn[
  \begin{@twocolumnfalse}
\noindent\LARGE{\textbf{Rheology of Cubic Blue Phases}}
\vspace{0.6cm}

\noindent\large{\textbf{Oliver Henrich $^{\ast}$\textit{$^{a,b}$}, Kevin Stratford \textit{$^{a}$}, Peter V. Coveney \textit{$^{b}$}, Michael E. Cates \textit{$^{c}$}, Davide Marenduzzo \textit{$^{c}$}}}\vspace{0.5cm}
%Please note that \ast indicates the corresponding author(s) but no footnote text is required. 

%\noindent\textit{\small{\textbf{Received Xth XXXXXXXXXX 20XX, Accepted Xth XXXXXXXXX 20XX\newline
%First published on the web Xth XXXXXXXXXX 200X}}}

\noindent \textbf{\small{DOI: 10.1039/c3sm50228g}}
\vspace{0.6cm}
%Please do not change this text.

\noindent \normalsize{We study the behaviour of cubic blue phases under shear flow 
via lattice Boltzmann simulations. We focus on the two experimentally observed
phases, Blue Phase I (BPI) and Blue Phase II (BPII). The disclination network of Blue Phase 
II continuously breaks and reforms under steady shear, leading to an oscillatory stress response in time.
For larger shear rates, the structure breaks up 
into a Grandjean texture with a 
cholesteric helix lying along the flow gradient direction.
Blue Phase I leads to a very different response. Here, oscillations are only possible for intermediate
shear rates -- 
very slow flow causes a transition of the initially 
ordered structure into an amorphous network 
with an apparent yield stress. 
Larger shear rates lead to another amorphous state with different structure of 
the defect network.
For even larger flow rates the same break-up into a Grandjean texture as for Blue Phase II is observed.
At the highest imposed flow rates both cubic blue phases adopt a flow-aligned nematic state.
Our results provide the first theoretical investigation of sheared blue phases in large systems, 
and are relevant to understanding the bulk rheology of these materials.}
\vspace{0.5cm}
 \end{@twocolumnfalse}
  ]

%\footnotetext{\dag~Electronic Supplementary Information (ESI) available: [details of any supplementary information available should be included here]. See DOI: 10.1039/b000000x/}

%Please use \dag to cite the ESI in the main text of the article.
%If you article does not have ESI please remove the \dag symbol from the title and the above footnotetext.

\footnotetext{\textit{$^{a}$~EPCC, School of Physics and Astronomy, University of Edinburgh,\\JCMB Kings Buildings, Mayfield Road, Edinburgh EH9 3JZ, UK; E-mail: ohenrich@epcc.ed.ac.uk}}
\footnotetext{\textit{$^{b}$~Centre for Computational Science, University College London,\\20 Gordon Street, London WC1H 0AJ, UK}}
\footnotetext{\textit{$^{c}$~SUPA, School of Physics and Astronomy, University of Edinburgh,\\JCMB Kings Buildings, Mayfield Road, Edinburgh EH9 3JZ, UK}}

%additional addresses can be cited as above using the lower-case letters, c, d, e... If all authors are from the same address, no letter is required

%\footnotetext{\ddag~Additional footnotes to the title and authors can be included \emph{e.g.}\ `Present address:' or `These authors contributed equally to this work' as above using the symbols: \ddag, \textsection, and \P. Please place the appropriate symbol next to the author's name and include a \texttt{\textbackslash footnotetext} entry in the correct place in the list.}

\section{Introduction}

Cholesterics are liquid crystals in which the local nematic director field shows spontaneous 
twist in thermodynamic equilibrium~\cite{deGennes}. 
The simplest manifestation is the standard cholesteric phase where the
director precesses around a single helical axis of fixed orientation.
For highly chiral systems, however,
the preferred configuration close 
to the isotropic boundary features twist around two perpendicular axes, as opposed to 
just one axis in the regular cholesteric state, and the corresponding deformation is 
denoted a ``double-twist cylinder''.
As it is topologically impossible to cover continuously 3D space with double-twist 
cylinders, defects arise. The resulting disclination lines (at which the
nematic director is undefined)
 organise into a variety of regular periodic lattices, 
giving rise to the so-called cubic blue phases (BPs)~\cite{Grebel:1984,Wright:1989}. 
There are two experimentally observed cubic blue phases, BPI and BPII (a third, BPIII, 
is thought to be amorphous~\cite{Henrich:2011a}).

BPs were long considered as purely of academic interest due to their very narrow 
range of stability. This view has changed since the creation of polymer-stabilised and other thermally 
stabilised BPs~\cite{Kikuchi:2002,Coles:2005}, which has opened up the 
possibility of novel applications.
During the last few years considerable progress has been achieved regarding the behaviour 
of BPs in confined geometries~\cite{Fukuda:2010a, Fukuda:2010b, Ravnik:2011b}, under 
external fields~\cite{Alexander:2008,Fukuda:2009,Henrich:2010a,Castles:2010,Tiribocchi:2011a}, 
and in the presence of colloidal particles~\cite{Ravnik:2011a}.
The kinetics of BP domain growth have been recently addressed~\cite{Henrich:2010b}. 
However, our understanding of their dynamical behaviour under flow remains
very limited. The aim of this work is to address this issue by studying,
for the first time, the response of large BP samples to a shear flow.

Flow response in cholesterics is both strongly non-Newtonian and highly anisotropic.
For example, if a standard cholesteric phase is subjected to a Poiseuille flow along
its helical axis, small pressure differences drive flow mainly through
``permeation'', as first investigated by Helfrich~\cite{Helfrich:1969}.
In the permeation mode the liquid crystal flows while leaving the director
field virtually unchanged, which leads to high dissipation and large
viscosities. 
Early experiments with cholesteric liquid crystals
showed that flow can also give rise to conformational transitions. \cite{Press:1978}.
Marenduzzo et al.~\cite{Marenduzzo:2006a,Marenduzzo:2006b} simulated 
shear and Poiseuille flow in cholesteric liquid crystals in the permeation mode, and 
showed the importance of the boundary conditions in determining the apparent viscosity of the fluid. 
They also found that a strong secondary flow appears.
Rey~\cite{Rey:1996a, Rey:1996b} theoretically studied shear in cholesterics oriented with the helix along 
the vorticity axis and found that, at low Ericksen number, travelling twist waves appear which 
lead to the rotation of the cholesteric helix. At higher forcing, the helix uncoils, creating a flow-induced nematic phase. 
However, this result was derived under the 
assumption that the molecules rotate only in flow-gradient plane whilst
the orientation of the cholesteric helix remained unchanged.
Rey also studied cholesterics subjected to both steady flow and low frequency
small amplitude oscillatory shear for different helix orientations
~\cite{Rey:2000, Rey:2002}. He found that splay/bend/twist deformations were
excited when the helix was aligned along the flow direction; splay/bend
deformation occurred when the helix was aligned along the velocity gradient;
but only twist deformations
appeared when the helix was aligned along the vorticity axis.

Dupuis et al.~\cite{Dupuis:2005} performed the first numerical investigation of BP rheology 
in Poiseuille flow, starting from equilibrium structures of BPI and BPII and a periodic 
array of doubly twisted cylinders.
Under small forcing, the network opposed the flow, giving rise to a significant 
increase in apparent viscosity.
Upon increasing the forcing they found clear evidence of shear thinning.
In the crossover region they predicted a novel oscillatory regime where the network 
continuously breaks and reforms as portions of the disclinations in the centre of the channel 
move to neighbouring cells and relink with the parts of the network left behind by the flow. 
The viscosity still decreases with forcing 
(the system shear thins) but much less than
for cholesterics in the permeation mode, which is in agreement with experiments
~\cite{Zapotocky:1999, Ramos:2002}.

Our work differs from these earlier efforts as it addresses homogeneous shear flows, 
and focuses on flow-induced reconstruction and nonequilibrium transition
between different blue phase networks, which appear at intermediate or
high shear. Our simulations employ Lees-Edwards boundary conditions, which
are naturally suited to address these regimes, and bulk as opposed
to boundary-dominated flow. 
Our simulations are large scale and parallel, so that we can study
significantly larger systems than previously possible, and our
results can in principle be compared with bulk rheology experiments. 

Our paper is organised as follows. 
In Section II, we describe the method we use and review
the hydrodynamic equations of motion which we aim to solve. In Section III, 
we report our numerical results, separating them into subsections referring to
Blue Phase I and Blue Phase II, and corresponding to low, intermediate, and
high shear. Finally, we draw our conclusions in Section IV.

\section{Model and Methods}

Our approach is based on the well-established Beris-Edwards model for hydrodynamics of
cholesteric liquid crystals \cite{Beris:1994}, which describes the ordered state 
in terms of a traceless, symmetric tensor order parameter ${\mathbf Q}({\mathbf r})$. 
In the uniaxial approximation, the order parameter is given by
$Q_{\alpha \beta}= q_s ( \hat{n}_\alpha \hat{n}_\beta - \frac{1}{3}\; \delta_{\alpha\beta})$
with $\hat{{\mathbf n}}$ the director field and $q_s$ the amplitude of nematic
order. More generally,
the largest eigenvalue of ${\mathbf Q}$, $0\le q_s\le\frac{2}{3}$
characterises the local degree of orientational order.
The thermodynamic properties of the liquid crystal are determined by a free energy
${\cal F}$, whose density $f$ consists of a bulk contribution $f_b$ and a gradient part $f_g$, as follows,
\begin{eqnarray}
f_b&=&\frac{A_0}{2}\left(1-\frac{\gamma}{3}\right) Q_{\alpha \beta}^2\nonumber\\
&-&\frac{A_0 \gamma}{3}Q_{\alpha \beta} Q_{\beta \gamma} Q_{\gamma \alpha}+\frac{A_0 \gamma}{4}(Q_{\alpha \beta}^2)^2,\nonumber\\
f_g&=&\frac{K}{2}(\varepsilon_{\alpha\gamma\delta} \partial_\gamma Q_{\delta\beta}+2 q_0 Q_{\alpha \beta})^2+\frac{K}{2}(\partial_\beta Q_{\alpha \beta})^2.\label{FE}
\end{eqnarray}
The first term contains a bulk-free energy constant $A_0$ and the temperature-related parameter $\gamma$ which controls the magnitude of order.
The second part quantifies the cost of elastic distortions, which is proportional to the elastic constant $K$;
we work for simplicity in the one-elastic constant approximation~\cite{deGennes}. The wavevector $q_0$ is equal to $2\pi/p_0$, where $p_0$ is the cholesteric pitch.
The actual periodicity of the BP structure, $p$, does not need to be equal to $p_0$.
Indeed, the ``redshift'' $r=p/p_0$ is adjusted during the equilibration phase of the 
simulation, to optimise the free energy density
before shearing begins -- this is done by following the procedure
previously described in~\cite{Alexander:2006}.

A thermodynamic state is specified by two dimensionless quantities: the reduced temperature 
\begin{equation}
\tau=\frac{27(1-\gamma/3)}{\gamma},
\end{equation}
which vanishes at the spinodal point of a nematic ($q_0=0$), 
and the reduced chirality 
\begin{equation}
\kappa=\sqrt{\frac{108 K q_0^2}{A_0 \gamma}},
\end{equation}
which measures the ratio of gradient to bulk free energy.

The dynamical evolution of the order parameter is given by the equation 
\begin{equation}
\left(\partial_t+ v_\alpha \partial_\alpha \right){\mathbf Q} - {\mathbf S}({\mathbf W},{\mathbf Q}) = \Gamma {\mathbf H}.
\label{op-eom}
\end{equation}
The first term on the left hand side of Eq.\ref{op-eom} is a material derivative, which describes the rate of change of a quantity advected by the flow.
The second term accounts for the rate of change due to local velocity gradients $W_{\alpha \beta}=\partial_\beta v_\alpha$,
and is explicitly given by~\cite{Beris:1994}
\begin{eqnarray}
{\mathbf S}({\mathbf W}, {\mathbf Q}) &=& (\xi {\mathbf A} + {\boldsymbol \Omega})({\mathbf Q}+\frac{\mathbf I}{3})\nonumber\\
& &\hspace*{-1.5cm}+ ({\mathbf Q}+\frac{\mathbf I}{3})(\xi {\mathbf A}  - {\boldsymbol \Omega})-2 \xi ({\mathbf Q}+\frac{\mathbf I}{3})
\mathrm{Tr}({\mathbf Q \mathbf W}),
\label{sw}
\end{eqnarray}
where $\mathrm{Tr}$ denotes the tensorial trace, while 
${\mathbf A}=({\mathbf W}+{\mathbf W}^T)/2$ and
${\boldsymbol \Omega}=({\mathbf W}-{\mathbf W}^T)/2$ are the symmetric and antisymmetric part of the velocity gradient, respectively. $\xi$ 
is a constant depending on the molecular details of the liquid crystal.
Flow alignment occurs if $\xi \cos{2\theta}=(3q_s)/(2+q_s)$ has a real solution for $\theta$, the 
so-called Leslie angle: we select this case by 
setting $\xi=0.7$ in our simulations.
${\mathbf H}$ is the molecular field, which is a functional derivative of $\cal F$ that respects the tracelessness of $\mathbf Q$:
\begin{equation}
{\bf H}=-\frac{\delta {\cal F}}{\delta {\bf Q}}+\frac{\bf I}{3}\,
\mathrm{Tr} \left(\frac{\delta {\cal F}}{\delta {\bf Q}}\right).
\label{molfield}
\end{equation}
The rotational diffusion constant $\Gamma$ in Eq.~\ref{op-eom} is proportional
to the inverse of the rotational viscosity $\gamma_1=2 q_s^2/\Gamma$
\cite{deGennes}.

The time evolution of the fluid density and velocity are respectively governed
by the continuity equation
$\partial_t \rho = -\partial_\alpha(\rho v_\alpha)$, and
the following Navier-Stokes equation:
\begin{eqnarray}
\partial_t v_\alpha +\rho \,v_\beta \partial_\beta v_\alpha
&=& \partial_\beta \Pi_{\alpha \beta}+ \eta\, \partial_\beta [ \partial_\alpha v_\beta +\; \partial_\beta v_\alpha].
\label{NSE}
\end{eqnarray}
This emerges from the Chapman-Enskog expansion
of the lattice Boltzmann (LB) equations 
that we solve numerically. A further term $\eta(1+3\frac{\partial P_0}{\partial\rho} )\partial_\mu v_\mu \delta_{\alpha \beta}$ that formally appears
in this expansion is negligible under the slow flows considered here
for which the fluid motion is almost incompressible~\cite{Denniston:2001}.
$\eta$ is an isotropic background viscosity which is set to $\eta=0.6666$ in LB units (these are discussed below, see~\cite{Henrich:2011a,Henrich:2010b}).
The thermodynamic stress tensor reads explicitly
\begin{eqnarray}
\Pi_{\alpha \beta}&=&P_0 \delta_{\alpha\beta}
-\xi H_{\alpha \gamma}\left(Q_{\gamma \beta} +\frac{1}{3} \delta_{\gamma \beta}\right)\nonumber\\
&-&\xi \left(Q_{\alpha \gamma} +\frac{1}{3} \delta_{\alpha \gamma}\right) H_{\gamma \beta} + Q_{\alpha \gamma}H_{\gamma \beta}-H_{\alpha \gamma} Q_{\gamma \beta} \nonumber\\
&+&2 \xi  \left(Q_{\alpha \beta} +\frac{1}{3} \delta_{\alpha \beta}\right) Q_{\gamma \nu} H_{\gamma \nu}
- \partial_\alpha Q_{\gamma \nu} \frac{\delta{\cal F}}{\delta \partial_{\beta} Q_{\gamma \nu}}\nonumber\\
\label{Pi}
\end{eqnarray}
and is responsible for strong non-Newtonian flow effects.
In the isotropic state ${\bf Q}\equiv 0$ and Eq.\ref{Pi} reduces to the
scalar pressure as would appear in Eq.~\ref{NSE} for a Newtonian fluid. 

We next define a 
dimensionless number that describes the deformation
of the director field under flow. This so-called Ericksen number
is given by 
\begin{equation}
Er=\frac{\eta v l}{K}
\end{equation}
with $\eta$ and $K$ defined previously, and 
$v$ and $l$ a typical velocity and length scale. 
In the present work $l=p_0/2=\pi/q_0$ was used as this is the approximate size of
the BP unit cell. Likewise, $v$ was taken to be the velocity difference
across one unit cell, i.e. $v=\dot{\gamma}\, l$. 
The elastic constants were $K=0.02$ (BPII) and $K=0.04$ (BPI), respectively.

The system of coupled partial differential equations~\ref{op-eom}
and~\ref{NSE} is solved by means of a hybrid method~\cite{Marenduzzo:2007}
which uses a combination of lattice Boltzmann and finite difference schemes.
(This is in contrast with some earlier methods using solely
LB~\cite{Denniston:2001, Denniston:2004}.)
The Navier-Stokes
equation is solved via the lattice Boltzmann approach, using a standard
three-dimensional model with 19 discrete velocities (D3Q19).
A regular lattice with spacing $\Delta x = \Delta y = \Delta z = 1$ is
used and the time step is $\Delta t = 1$ in lattice units.
Coupling to the thermodynamic sector is via a
local body force computed as the divergence of the thermodynamic
stress Eq.~\ref{Pi}; the resulting velocity field is used in the computation
of the time evolution of $\mathbf{Q}$ via a standard finite difference
method using the same grid and the same time step as the LB. The system
is periodic in all three coordinate directions. Sliding
boundary conditions, or Lees-Edwards planes ~\cite{Wagner:2002}, are explicitly implemented
in both hydrodynamic and thermodynamic sectors to impose shear on the
system. 
The same method with Lees-Edwards planes
has been used successfully to study binary mixtures, particle suspensions~\cite{MacMeccan:2009, Aidun:2010}, 
emulsions~\cite{Frijters:2012} and smectic liquid crystals~\cite{Henrich:2012a}.
A more detailed discussion of our use of Lees-Edwards boundary conditions can be found in the Appendix.

In the following we report results of simulations of the bulk flow behaviour of the cubic Blue 
Phases BPI and BPII.
Typical runtimes for system size $L_x\times L_y\times L_z=128^3$ on 512 processes were in the region of 18 to 24 hours.  
The timestep and lattice spacing in lattice Boltzmann units (LBU) can be mapped
approximately to $\sim 1 {\rm ns}$ and $\sim 10{\rm nm}$ in SI units, respectively. The LB unit of stress
is equal to about $10^8$~Pa. Further details about the conversion 
from LBU to SI units can be found elsewhere~\cite{Henrich:2011a,Henrich:2010b}.
In what follows, $x$, $y$ and $z$ denote respectively the velocity, velocity
gradient and vorticity direction; $\Pi_{xy}$ is therefore the shear stress.

\section{Results and Discussion}

For typical simulations reported in this work, 
we chose as initial conditions thermodynamic states that are 
well inside the equilibrium region of the individual blue phase, and far away 
from the cholesteric-isotropic transition. Thus, temperature and chirality were 
$\tau=-0.5, \kappa=1.0$ in case of BPI and $\tau=-0.5, \kappa=2.0$ for BPII, respectively.
For these parameters the total free energy density $f$ remained always negative at all flow rates
simulated. 
Since by Eq.~\ref{FE} $f=0$ for an isotropic phase with ${\mathbf Q}\equiv 0$, 
this means that our system, which is never far from equilibrium locally even under flow, always remains in a liquid crystalline state. 
We also performed selected simulations on metastable states at higher and lower temperatures, 
and at different chiralities, but did not find any significant differences in the 
general flow behaviour from that described below.
The only quantitative difference we found was that, for thermodynamic states that are closer to the 
phase boundary than the one we focused on (and describe below), the critical shear
rate at which the disclination 
network broke up into chiral nematic states was lower. This is expected as these states have on average 
higher free energy densities and smaller order parameters than those we focus on in what follows, 
and cannot therefore withstand the same external forces before breaking down.

As usual in BP simulation studies~\cite{Henrich:2011a,Henrich:2010b}, we initialised our runs with 
analytical solutions that minimise the free energy functional Eq.\ref{FE} in the high-chirality limit 
and equilibrated these configuration for 5000 LB timesteps before we started the shear flow. 
During the equilibration sequence the optimal redshift $r$ was calculated and applied at every timestep.
After equilibration the redshift was kept constant throughout the rest of the simulation with shear flow (as 
it is no longer justified to optimise the free energy in a nonequilibrium scenario).
We chose a pitch length of 32 LBU for BPII and 64 LBU in case of BPI, and we considered in both cases 
4 unit cells along each coordinate direction, for a total of 64 unit cells in our simulation box.
Runs with higher resolution confirmed that this choice was adequate to track  
all kinematic details of the blue phase networks in shear flow. This includes
reconstruction of the unit cell not accessible in simulation with only
few cells~\cite{Dupuis:2005}.

Simple shear flow was imposed by means of the Lees-Edwards boundary
conditions with
the top (bottom) part of the system flowing in the positive (negative) $x$-direction and the 
velocity gradient along the $y$-direction.
The shear rates were varied over more than three orders of magnitude from about 
$\gd=2.44\times \e{-6}$ to $3.75\times\e{-3}$ LBU.
For clarity we classify various flow regimes, namely three in the case of Blue Phase II and five in the 
case of Blue Phase I. These regimes comprise those with periodically recurring conformations and oscillatory stress response (PRC), 
amorphous networks (AN), as well as those featuring a Grandjean texture (GJ) which 
may be partly frustrated (FGJ) and a flow-aligned nematic state (FN). 
GJ and FN occur in both blue phases as steady states regardless of the initial state.

These regimes are characterised by the following approximate flow rates: 

\begin{table}[htpb]
\begin{tabular}{|c|c|c|}
\hline
Name & $\gd$ & $Er$ \\
\hline
BPI-1 (AN) & $ \lesssim 1.95\e{-5}$ & $\lesssim 0.17$ \\
BPI-2 (PRC) & $3.91\e{-5} \dots 2.34\e{-4}$ & $0.33 \dots 2$ \\
BPI-3 (AN) & $3.13\e{-4}\dots 4.69\e{-4}$ & $2.67 \dots 4$ \\
BPI-4 (GJ, FGJ) & $5.47\e{-4} \dots 1.88\e{-3}$ & $4.67 \dots 16$ \\
BPI-5 (FN) & $\gtrsim 2.5\e{-3}$ & $ \gtrsim 21.33$ \\
\hline
BPII-1 (PRC) & $\lesssim 3.91\e{-4}$ & $ \lesssim 3.33$\\
BPII-2 (GJ) & $4.69\dots 1.25\e{-3}$ & $4 \dots 10.67$ \\
BPII-3 (FN) &$\gtrsim 1.88\e{-3}$ & $ \gtrsim 16$ \\ 
\hline
\end{tabular}
\label{table2}
\end{table}

As is standard \cite{Henrich:2010b,Henrich:2012b} disclination lines are represented by plotting an isosurface of the scalar order parameter $q_s$. 
Typical choices are $q_s=0.17$ for BPI and $q_s=0.15$ for BPII.
(Note that $q_s$ is small but non-zero at the disclination core. 
The director is undefined there because the largest and second largest eigenvalues of ${\mathbf Q}$ coincide.)

\subsection{Blue Phase II}

We start our discussion with BPII as its flow behaviour is somewhat simpler 
than that of BPI. BPII has simple cubic symmetry and the disclination lines 
intersect and form a characteristic network of nodes.
Ahead of the more detailed discussion we provide first a general overview of the 
flow behaviour at all applied shear rates.

\begin{figure}[htpb]
\includegraphics[width=0.495\textwidth]{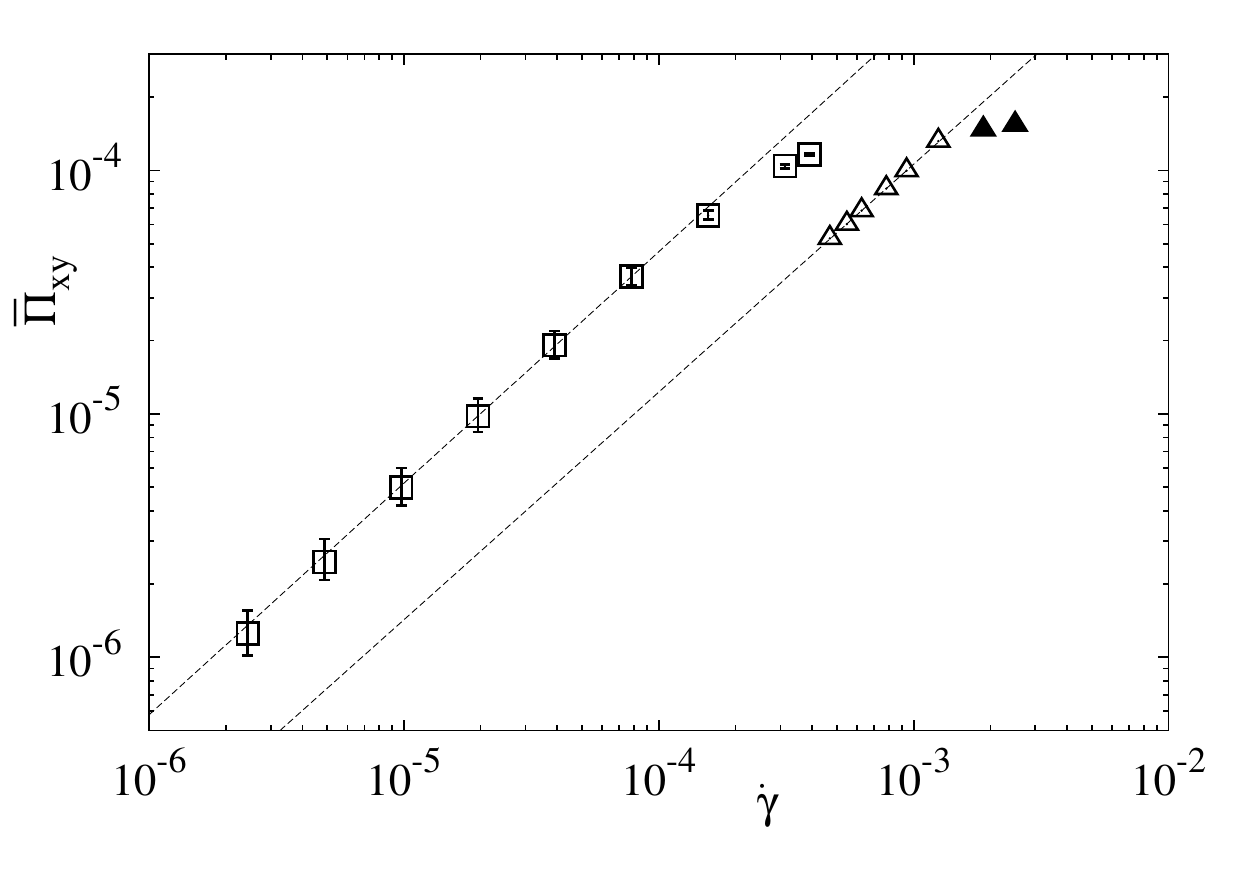}
\caption{
Flowcurve $\bar{\Pi}_{xy}(\gd)$ of BPII. The open squares show the average shear stress in the regime with periodically recurring conformations,
whereas the error bars indicate the maximum and minimum stresses that occur during one cycle. Above a critical flow rate the network  
breaks up into a Grandjean texture (open triangles) or a flow-aligned nematic state at even higher flow rates (solid triangles).
The two dashed lines represent fits to the data with $\bar{\Pi}_{xy}(\gd)\propto \gd^{\;0.95}$ (left) and $\bar{\Pi}_{xy}(\gd)\propto \gd^{\;0.94}$ (right), respectively.
}
\label{bp2-flowcurve}
\end{figure}

\begin{figure}[htpb]
\includegraphics[width=0.495\textwidth]{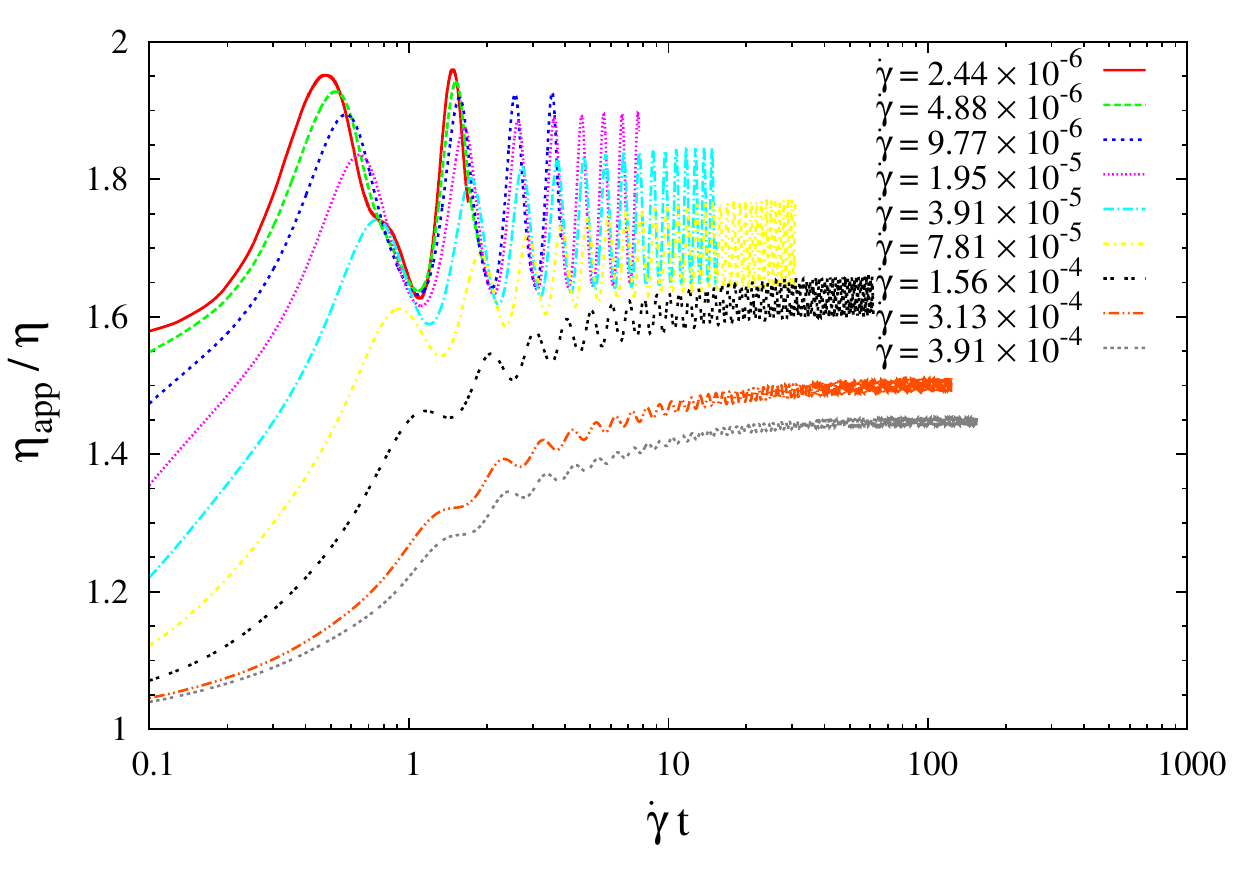}\\
\includegraphics[width=0.495\textwidth]{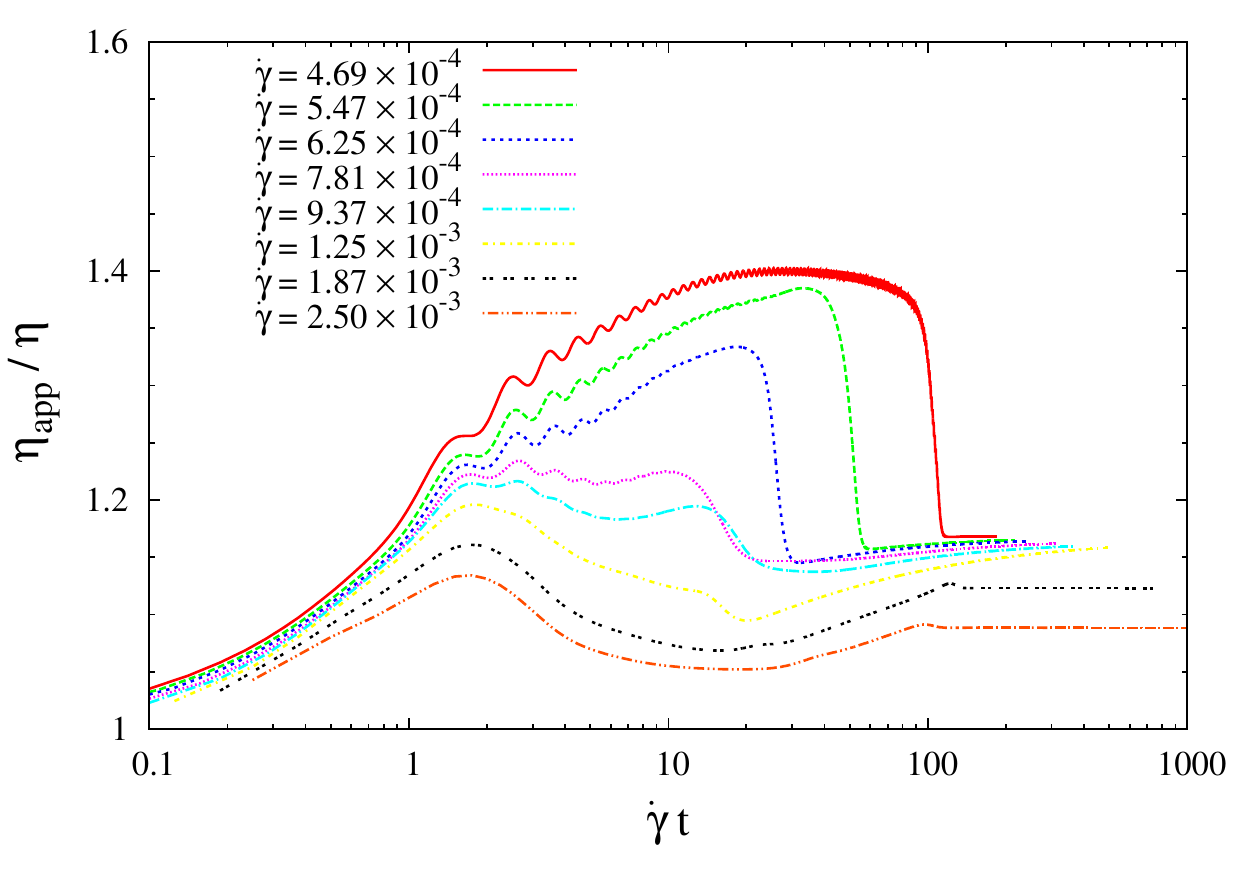}\\
\caption{
Apparent viscosity $\eta_{app}=\Pi_{xy}/\gd + \eta$ 
of BPII versus strain $\gamma = \gd\cdot t$  
normalised to the background viscosity $\eta$.
The pictures show the regimes BPII-1 (periodically recurring conformations, top), BPII-2 (Grandjean texture, bottom) and BPII-3 (flow-aligned nematic, bottom). 
The value $\eta_{app}/\eta=1$ corresponds to Newtonian flow without any additional contribution of the liquid crystal, i.e. totally shear-thinned state.
} 
\label{bp2-appvisc}
\end{figure}

\begin{figure}[htpb]
\center
\includegraphics[width=0.35\textwidth]{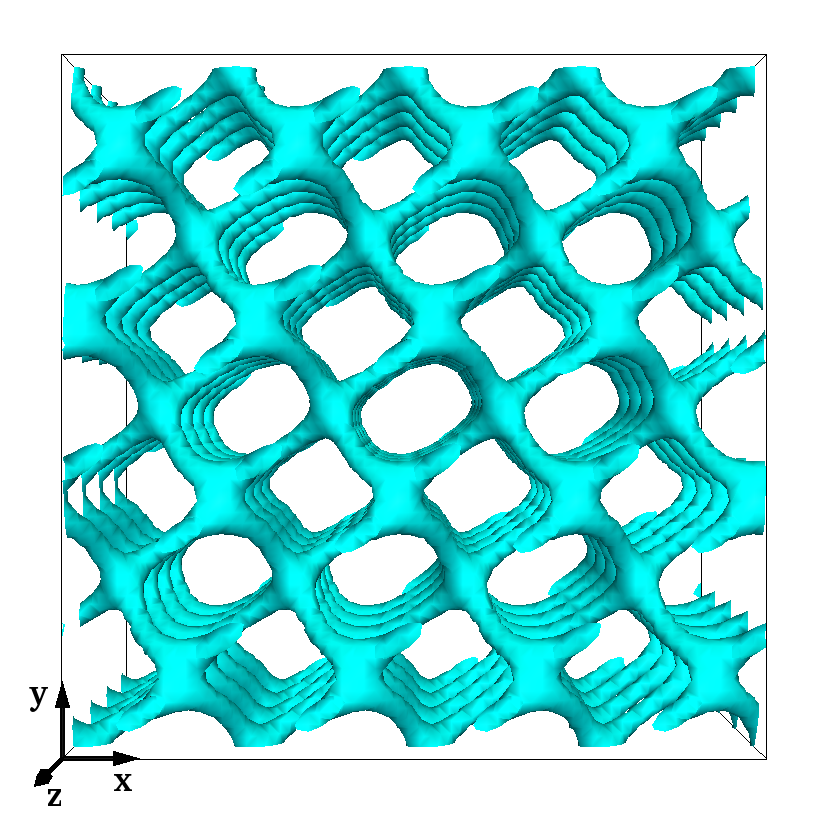}
\includegraphics[width=0.35\textwidth]{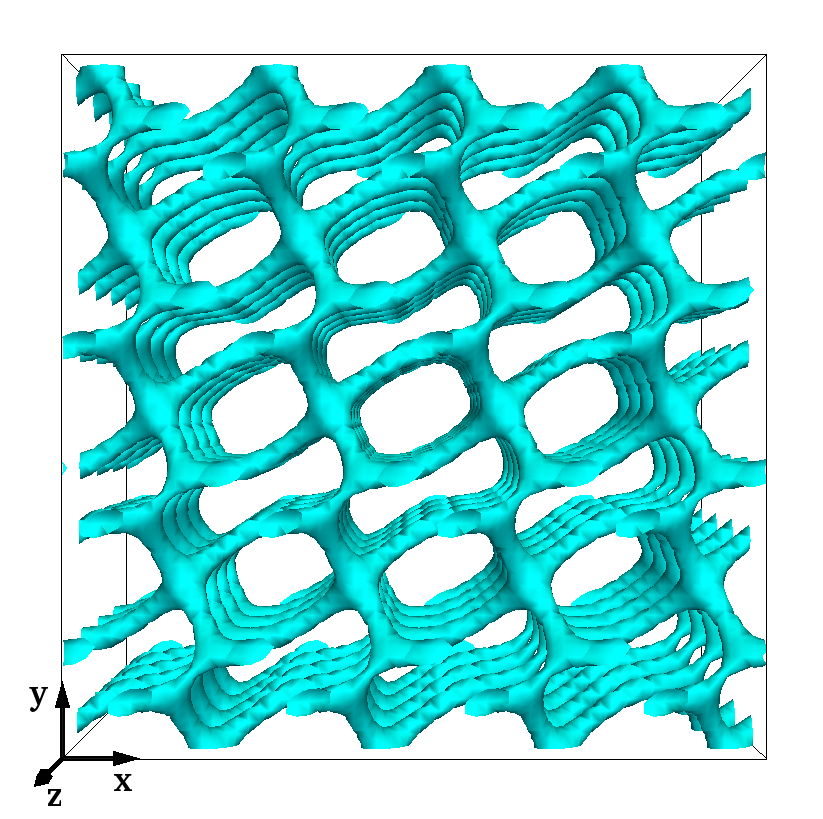}
\includegraphics[width=0.35\textwidth]{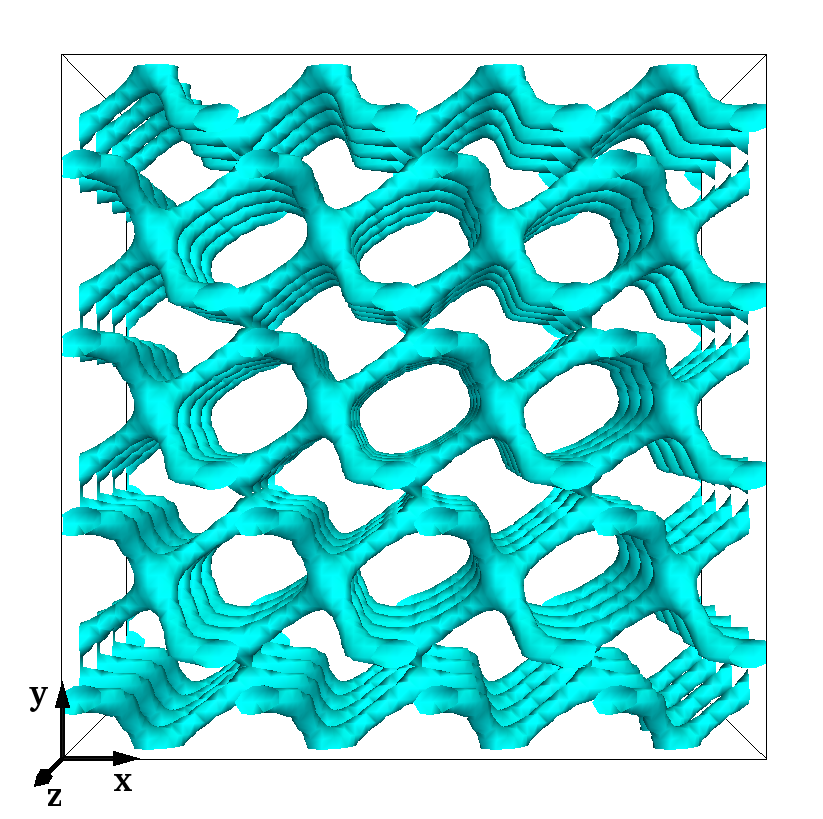}
\caption{Disclination network of BPII in shear flow: 
The pictures show a typical sequence of snapshots in the steady state 
at $\gd=1.56\e{-4}$ and time steps $t=1.60, 1.64,1.65\e{5}$. The velocity 
gradient is oriented along the vertical direction (y), whereas the 
horizontal direction (x) is the flow direction. Lees-Edwards boundary 
conditions have been imposed in such a way that the network moves to the 
right in the upper half and to the left in the lower half of the picture.}
\label{bp2-1-disc}
\end{figure}

Fig. \ref{bp2-flowcurve} shows a flow curve, defined as time averages of 
the shear stress $\bar{\Pi}_{xy}$ as a function of shear rate $\gd$.
Time averages were taken as average over one stress cycle in the steady state.
Where no oscillations occurred we used the final value at the end of the run.
For the lowest up to intermediate shear rates, the regime we refer to as BPII-1, a 
power law fit $\bar{\Pi}_{xy}=a \gd^b$ with $a=0.30, b=0.95$ describes 
the data to a very good approximation.
This holds true even in regime BPII-2, where the network breaks up and 
the liquid crystal adopts a Grandjean structure with the helical axis oriented along the
flow gradient direction ($y$). The coefficients of the power law fit are
here $a=0.68$ and $b=0.94$. Hence, for the range of shear rates which we have explored here, 
the degree of shear-thinning is remarkably small in BPII.

Fig. \ref{bp2-appvisc} shows the ratio between the apparent viscosity, 
defined as $\eta_{app}=\langle \Pi_{xy} \rangle/\gd +\eta$, and the background viscosity $\eta$ 
over total strain.
A numerical value of $\eta_{app}/\eta=1$ corresponds to a fully Newtonian flow,
without any additional stress contribution from the liquid crystal.
The top picture shows data for regime BPII-1, where the periodic break-up and reconnecting of the network 
in shear flow causes sinusoidal oscillations in $\eta_{app}$. As noted in the Appendix, the
absence of a permeation mode at very low shear rates may be due to our
choice of boundary conditions. We refer the reader to the Appendix for a detailed discussion.

We observe another configuration, lying between those with periodically recurring conformations 
and the flow-aligned nematic state at shear rates $4.67\e{-4}\lesssim\gd\lesssim1.25\e{-3}$ ($4 \lesssim {\it Er} \lesssim 10.67$). 
In this interval the network breaks up completely into a simple cholesteric liquid crystal, and oscillations in the stress signal are absent.
The helical axis lies along the flow gradient direction (the so-called Grandjean texture~\cite{deGennes})
and the liquid crystal is flowing in ``nematic planes'' with the director predominantly oriented in flow-vorticity plane and having only
a small angle with the flow direction.
The picture at the bottom of Fig.~\ref{bp2-appvisc} shows data for this regime, to which we refer as BPII-2. The 
curves for the two largest flow rates are for BPII-3, the flow-aligned nematic state. 
We will address these regimes in section \ref{gj-fan}, as they are common to both cubic blue phases BPI and BPII.

The travelling helical wave which was identified in~\cite{Rey:1996a,Rey:1996b} when shearing a cholesteric helix along
the vorticity axis turns out to be a metastable flow state as it features larger free energies, higher stresses and dissipation due 
to the tumbling motion of the director field, which are completely absent in the Grandjean texture.
For even higher shear rates the system undergoes a transition to a flow-aligned nematic state.

In the next sections we investigate the BPII-1 flow regime in more detail 
by looking at the kinetics of the disclination network.

\subsubsection{Regime BPII-1: low and intermediate shear rates }

\begin{figure}[htbp]
\includegraphics[width=0.495\textwidth]{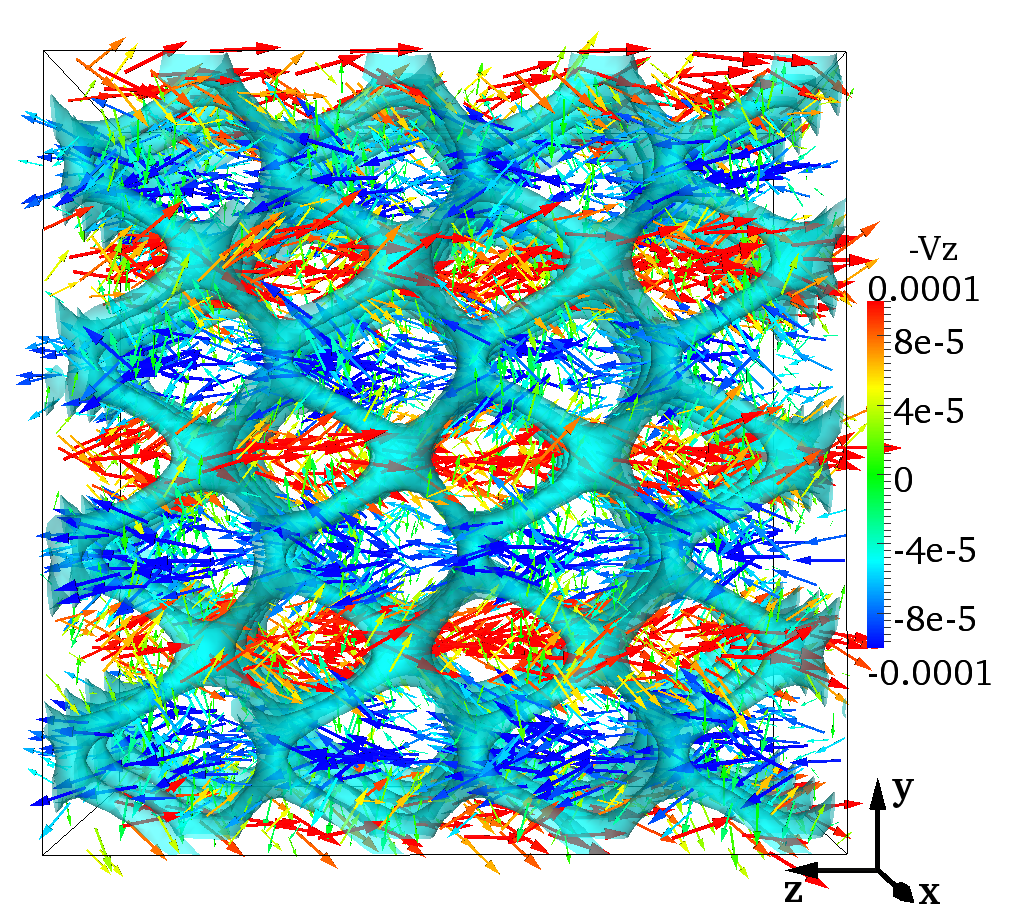}
\includegraphics[width=0.495\textwidth]{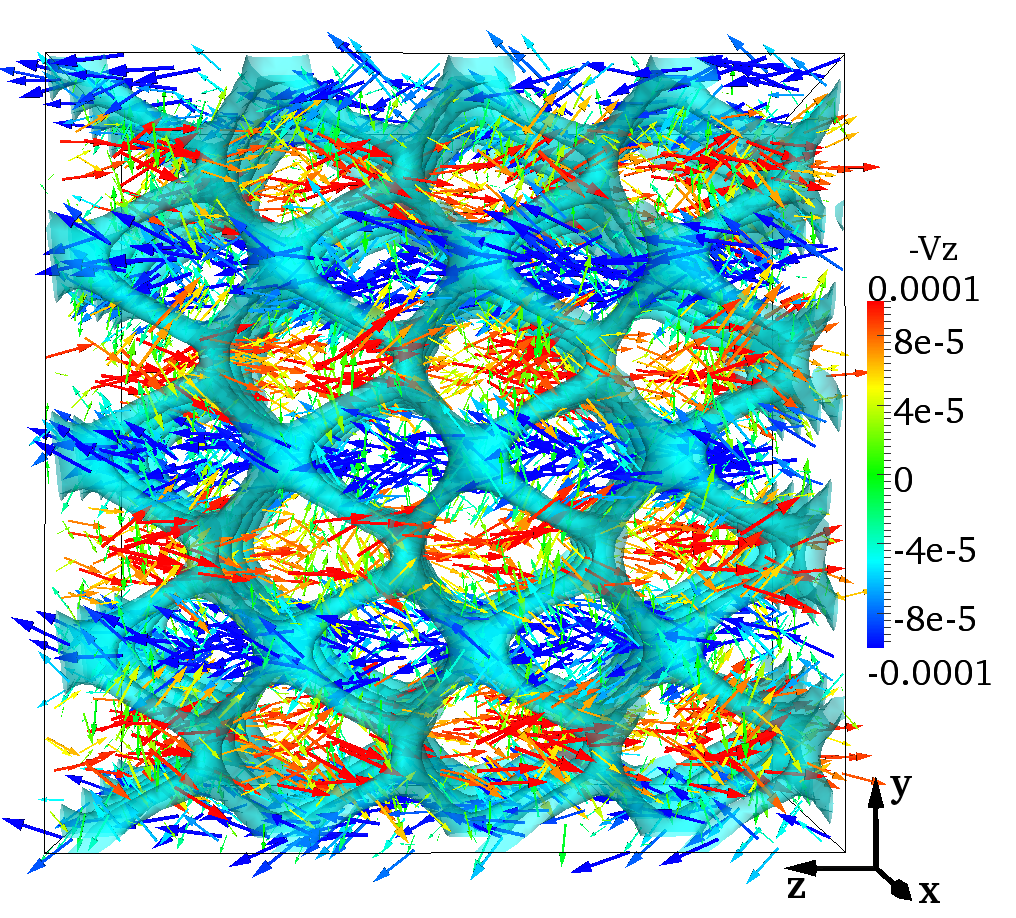}
\caption{Velocity patterns and disclination network in BPII for positive (left) and negative (right) helicity of the 
underlying cholesteric helix: The pictures show velocity vectors $(0,v_y,v_z)$.
The view is along the x-direction. This means the velocity ${\mathbf v}$ 
has been projected onto a plane perpendicular to the flow direction. 
The vertical and horizontal direction are the gradient and vorticity direction, respectively.
The colour code gives the magnitude and sign of the component in vorticity direction.
The snapshot shows a typical frame during a periodically recurring sequence.
The network on the left with right-handed helicity ($q_0>0$ in Eq.~\ref{FE}) travels rightwards, whereas the one one the right,
with reversed helicity, moves leftward. 
The periodicity of the motion along the vorticity direction is six times longer than the time 
it takes the network to reconnect along the flow direction.}
\label{bp2-1-velo}
\end{figure}

Fig. \ref{bp2-1-disc} shows the disclination network in shear flow as
it undergoes homogeneous shearing in the BPII-1 flow regime. 
The disclination lines break up and 
reconnect further downstream, forming a periodically recurring pattern. The
period needed for a pattern to break up and 
reform along the flow direction is $\tau_F = 1/\gd$.

The general appearance of the flowing network is, apart from the homogeneous distortion,
very close to that of the quiescent blue phase at equilibrium. This holds for all
shear rates in the BPII-1 regime.

Interestingly, while being displaced with the flow the entire network moves 
along the vorticity direction ($z$) as well.  A similar behaviour has been recently 
observed for blue phases in shear flow in confined geometries ~\cite{Henrich:2012b}.
In contrast to the stick-slip motion that has been reported there,
in this case the movement is steady. 
The vorticity motion occurs in such a way that the positions of breakup and reconnection 
point in the network, visible in Fig. \ref{bp2-1-disc}, are slightly offset and allow the network 
to travel along the $z$-direction. The periodicity of the motion along the vorticity direction
is $\tau_V=6\tau_F$, 
i.e. it takes a displacement of six unit cells along $x$ (the flow direction) 
for the network to move one unit cell along $z$ (the vorticity direction).

Fig. \ref{bp2-1-velo} shows the disclination network in the middle of a breakup-reconnection
cycle, with superimposed velocity vectors.
The plot shows the secondary velocity components, obtained by projecting the
velocity onto a plane perpendicular to the flow direction.
This gets rid of the dominating velocity component along the flow, $v_x$, and 
allows to visualise the patterns in the 
two much smaller components $v_y$ and $v_z$.
The magnitude of the secondary components is typically in the range of a 
few percent of the primary flow component, for the shear rates 
and system sizes simulated here.

Characteristic bands are visible in Fig. \ref{bp2-1-velo}, which are 
oriented along the vorticity direction. The direction of the flow
in each of the bands depends on the helicity of the underlying cholesteric
liquid crystal (i.e. on the sign of $q_0$ in Eq.~\ref{FE}), and the velocity
patterns for left-handed and right-handed BPs are one the mirror image of
the other. 
Further quantitative evidence for a direct link between the sense of motion 
and the helicity can be gained by time-averaging over individual cycles.

Table~\ref{tab1} in the Appendix 
gives minima, maxima, averages and standard deviations 
of the velocity components.
All values for the two runs with inverted helicity are identical apart 
from a change of sign in the $z$-components.
There is only a slight imbalance in magnitude 
between the maximum and minimum velocities 
along the $z$-direction. 
Therefore the averaged flow along the vorticity
direction is very small, and much smaller than
the local flow velocity. This suggests that the
flow of the network along the vorticity 
direction is permeative. This claim is further supported by the characteristic 
velocity bands in the secondary flow, visible in Fig.~\ref{bp2-1-velo}. 
The secondary flow in vorticity direction takes clearly place in positive {\em and} 
negative z-direction, whereas the movement of the network as a whole is only in one direction.

\subsection{Blue Phase I}

BPI has body-centred cubic symmetry and, unlike BPII, the disclination lines
characterising its equilibrium structure are well separated and do not 
intersect. (The quiescent state resembles the first frame in Fig.~\ref{bp1-1-disc}
below.)
We believe that this topological difference is responsible for most of
the differences between BPI and BPII regarding their rheological response. 
We present again key aspects in an overview before we address specific features 
in more detail. 

\begin{figure}[htpb]
\includegraphics[width=0.495\textwidth]{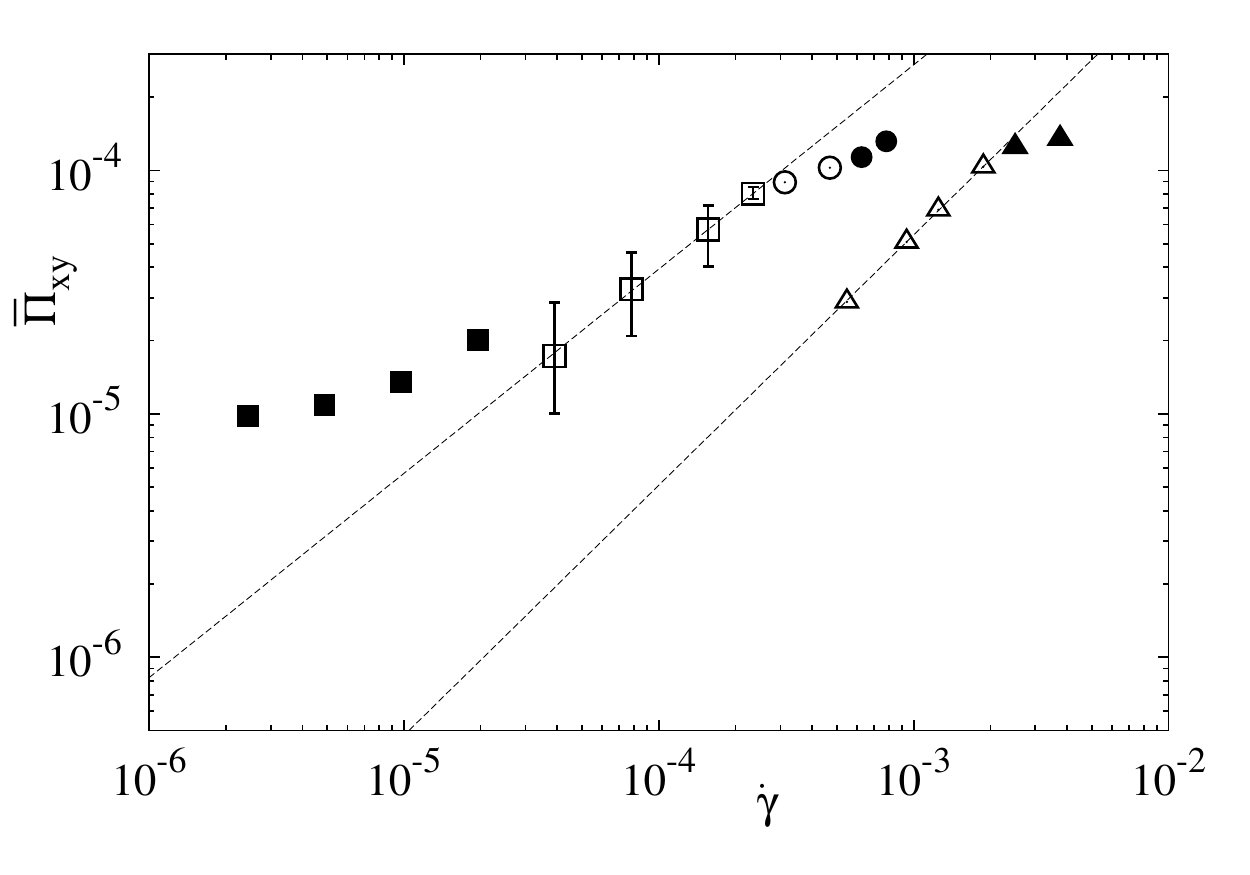}\\
\caption{
Flow curve $\bar{\Pi}_{xy}(\gd)$: 
Depending on the configuration in steady shear flow five different regimes 
can be distinguished: amorphous BP network with apparent yield stress (BPI-1, solid squares); 
steady flow with periodically recurring patterns (BPI-2, open squares); 
break-up of the disclination network into an amorphous state at $Er\simeq O(1)$ (BPI-3, open circles);
cholesteric helix with axis in gradient direction, also referred to
as Grandjean texture (BPI-4, open triangles), as well as frustrated metastable 
Grandjean textures (BPI-4, solid circles) that can transform into 
a regular Grandjean texture upon increasing the flow rate;
and flow-aligned nematic state (BPI-5, solid triangles). 
The error bars represent the minimum and maximum shear stress 
during one cycle in regime BPI-2.
The two dashed lines show fits to the data in regime BPI-2 with 
$\bar{\Pi}_{xy}(\gd)\propto \gd^{\;0.84}$ (left) 
and in regime BPI-4 with 
$\bar{\Pi}_{xy}(\gd)\propto \gd^{\;1.02}$ (right), respectively.}
\label{bp1-flowcurve}
\end{figure}

\begin{figure}[htpb]
\includegraphics[width=0.45\textwidth]{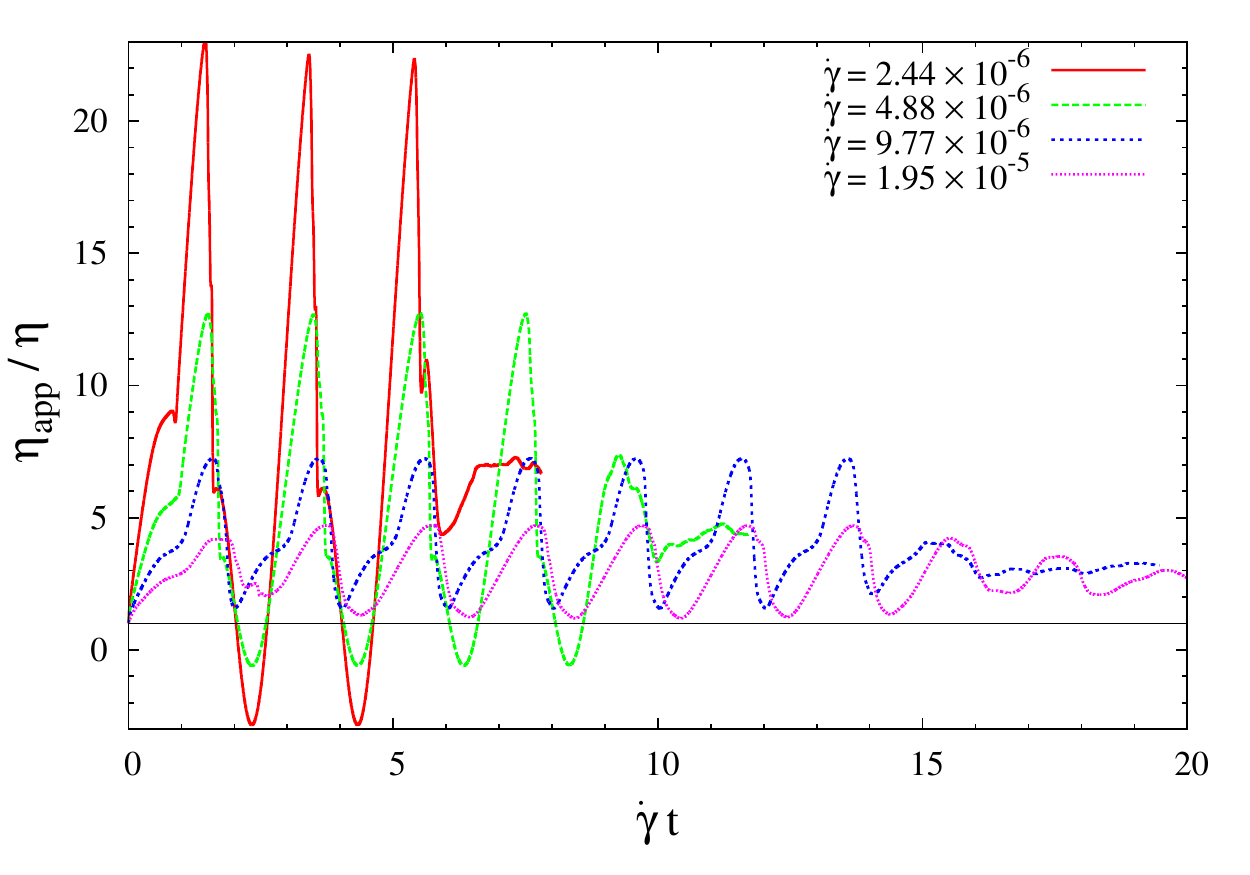}\\
\includegraphics[width=0.45\textwidth]{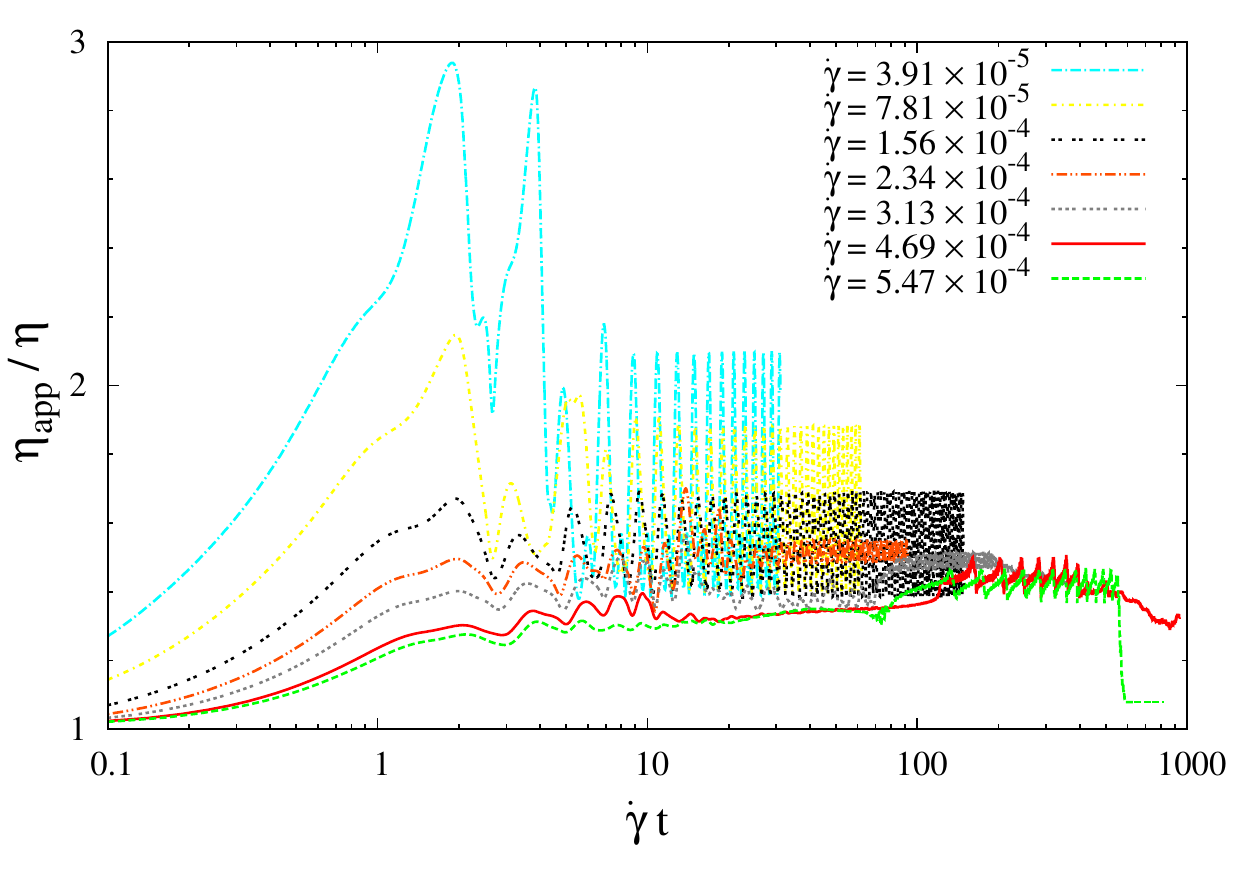}\\
\includegraphics[width=0.45\textwidth]{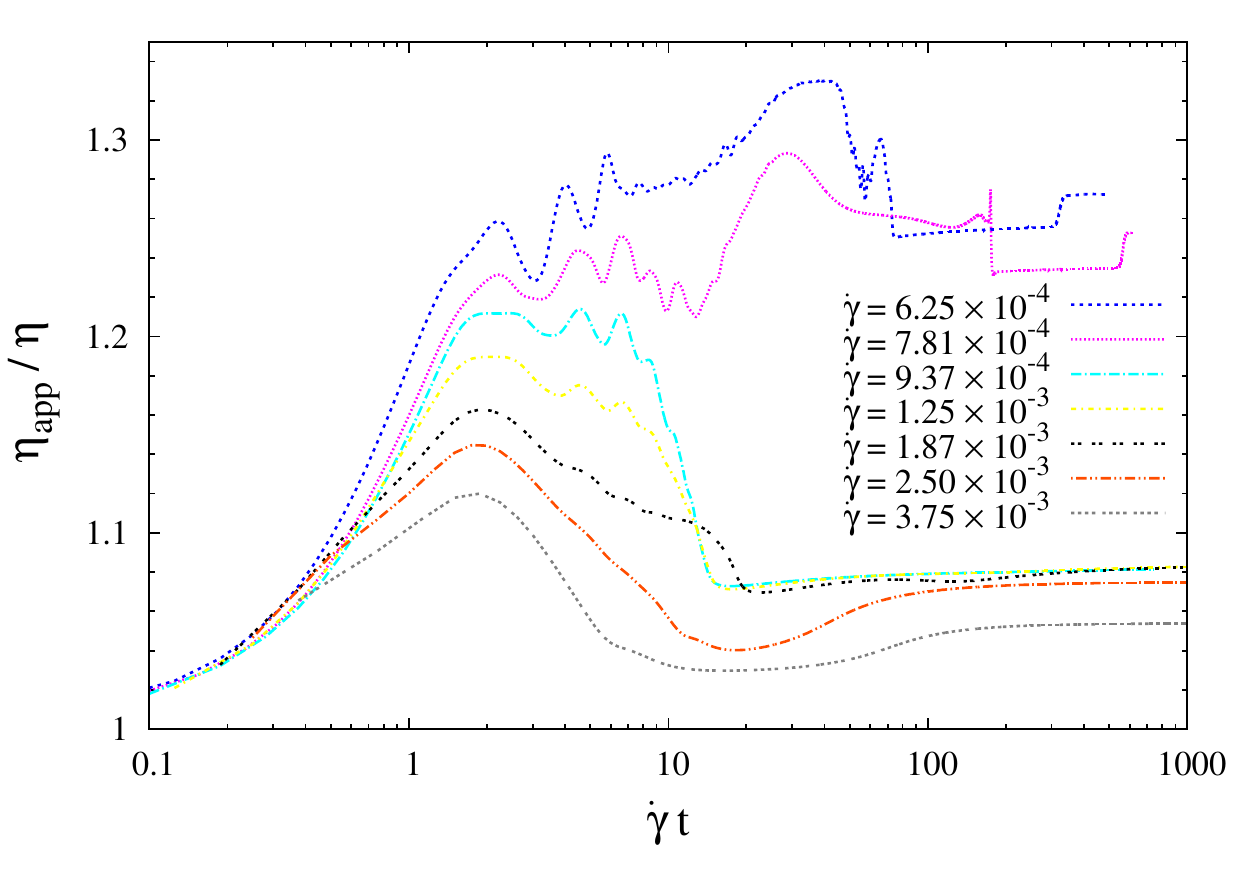}
\caption{
Apparent viscosity $\eta_{app}=\langle \Pi_{xy}\rangle/\dot{\gamma} + \eta$ of BPI versus strain normalised to the background viscosity $\eta$: 
The top picture shows regime BPI-1 (amorphous network) where
the BP network transforms into an amorphous network featuring a yield stress.
The centre picture shows regimes BPI-2 (periodically recurring conformations) and BPI-3 (amorphous network), 
whereas the picture at the bottom
depicts data of the (partly frustrated) Grandjean texture in regime BPI-4 
and the flow-aligned nematic state, labelled BPI-5 in this context. 
Note that the $\eta_{app}/\eta$ is sometimes negative for the lowest 
flow rates as the (negative) contribution of the elastic forces
of the network dominates the total stress.
}
\label{bp1-appvisc}
\end{figure}

Fig. \ref{bp1-flowcurve} shows the flow curve of BPI for the same range of
shear rates as in Fig.~\ref{bp2-flowcurve} and Fig. \ref{bp1-appvisc} depicts 
the apparent viscosity ratio $\eta_{app}/\eta$ as a function of total strain.

For intermediate (but not too low) shear rates
($3.91\e{-5}\lesssim\gd\lesssim 2.34\e{-4}; 0.33\lesssim {\it Er} \lesssim 2$), we once more 
find regular oscillations in the network morphology and in the stress
response versus time. We label this regime BPI-2. 
The centre picture of Fig.~\ref{bp1-appvisc} shows the corresponding data for the
apparent viscosity ratio versus strain.

The BPI-2 regime now does not persist to arbitrarily low shear: there the
oscillations become irregular and seemingly chaotic, and after a few
cycles the disclination pattern becomes amorphous.
For these low shear rates 
($2.4\e{-6}\lesssim\gd\lesssim 1.95\e{-4}; 0.02\lesssim {\it Er} \lesssim 0.17$) 
the flow curve in Fig. \ref{bp1-flowcurve}
develops a horizontal yield stress branch. 
We refer to this flow regime as BPI-1.

The BPI-2 regime is also unstable if the shear is increased {\em above} 
a critical value, 
which is in the region of Ericksen numbers $Er \simeq O(1)$.
This is a regime 
($3.13\e{-4}\lesssim\gd\lesssim 4.69\e{-4}; 2.67\lesssim {\it Er} \lesssim 4$) 
we labelled BPI-3. It is characterised by the balance between viscous and elastic forces.
This has an influence on the stability of the disclination network as it 
seems to become increasingly sensitive to small perturbations in the flow field
which couple back to the order parameter dynamics and vice versa. This phenomenon forms 
a possible route towards an instance of ``rheochaos'' ~\cite{rheochaos,Cates:2002}.

Finally, for shear rates such that
$5.47\e{-4}\lesssim\gd\lesssim1.88\e{-3}$ ($4.67\lesssim {\it Er} \lesssim 16$) 
and before the break-up into a flow-aligned nematic occurs 
(BPI-5, $2.5\e{-3}\lesssim\gd; 21.33\lesssim {\it Er}$) 
we observe the same cholesteric Grandjean configuration with the 
helical axis along the gradient direction as for BPII-2. 
For clarity we refer to this regime as BPI-4. It will be discussed 
in a separate section \ref{gj-fan}. 

\subsubsection{Regime BPI-1: low shear rates }

\begin{figure}[htpb]
\includegraphics[width=0.495\textwidth]{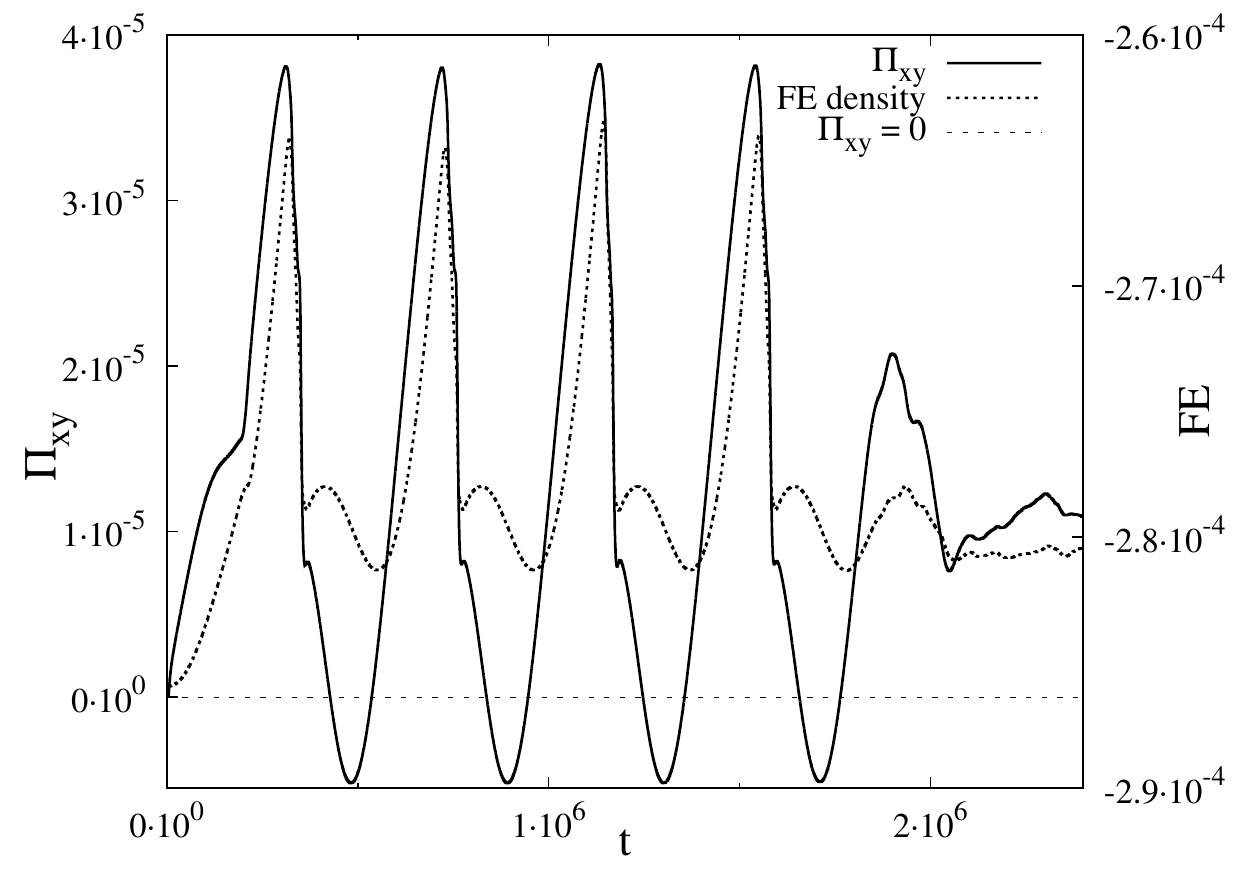}
\caption{Average thermodynamic shear stress and free energy density of 
BPI at shear rate $\gd=4.88\e{-5}, Er=0.08$: 
The negative branch in the stress is related to a local maximum 
and a following local minimum in average free energy density. 
This creates unstable conditions that lead to reconstruction of the 
defect lattice into an amorphous network state which 
seemingly features a yield stress.}
\label{bp1-fe-yield}
\end{figure}

The rheological response of BPI at low shear rate, $\gd\lesssim1.95\e{-5}$
(${\it Er} \lesssim 0.33$), 
is strikingly different from that of BPII and appears to show a
yield stress. An explanation for this behaviour can be found by 
looking more closely at the average shear stress (where the
background viscosity contribution has been subtracted) and 
free energy density as a function of time, as shown in Fig. \ref{bp1-fe-yield}.
When the quiescent and equilibrated BPI network begins to flow
the shear stress increases steeply and goes through a maximum.
Shortly after it goes negative 
to become positive again later on, forming a complete cycle.
The amplitude of the excursions in these stress oscillations, and
the presence of a part of the period where the thermodynamic
contribution to the stress is negative, suggest that the
BPI network is subject to large forces. Seemingly, these eventually
cause the flow-induced collapse to an amorphous network state with
an apparent yield stress. We should underscore that this
low shear regime might depend on our choice of boundary
conditions. In practice they are equivalent to 
having infinitely distant walls to which the network is anchored. 
See Appendix for a full discussion of our boundary conditions.

\begin{figure}[htpb]
\includegraphics[width=0.495\textwidth]{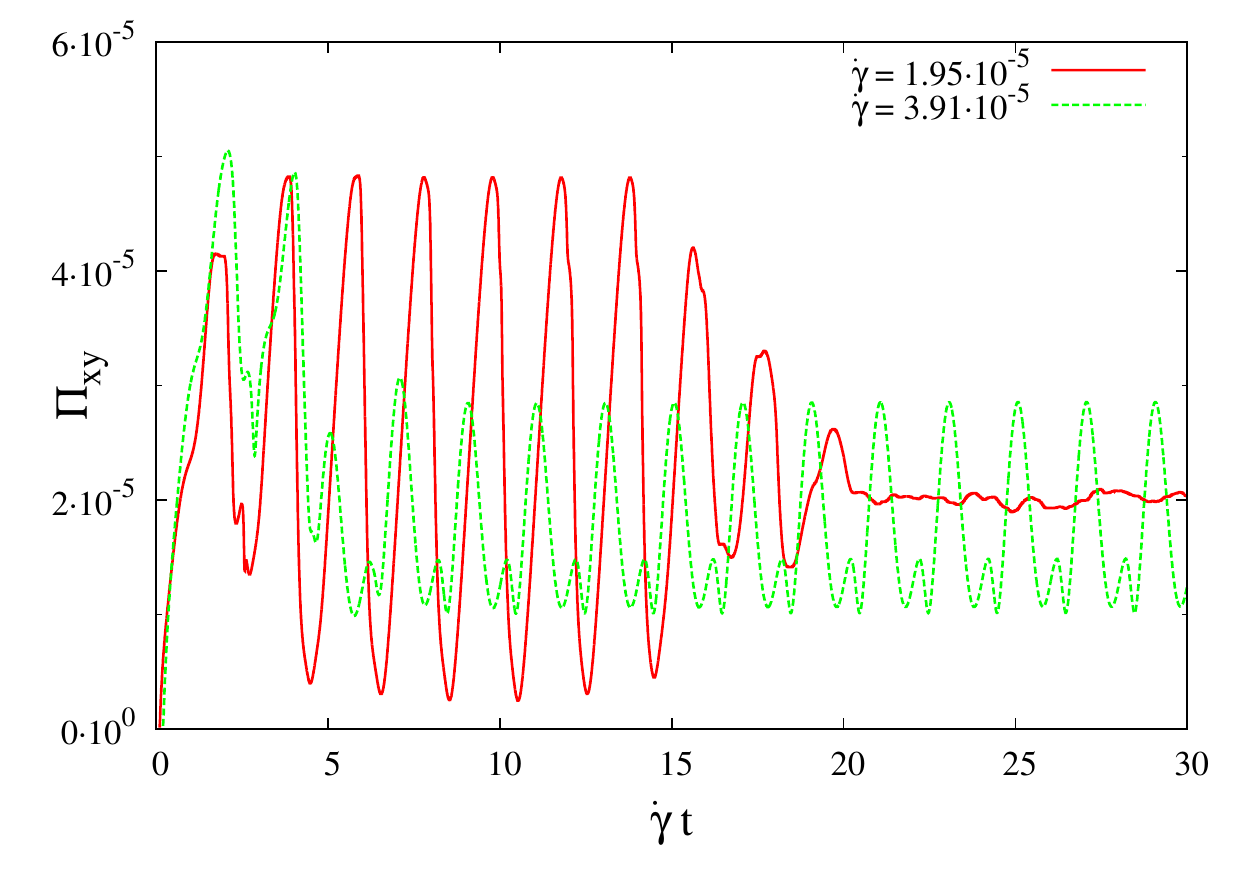}
\caption{Thermodynamic shear stress $\Pi_{xy}$ versus strain near the 
transition between regime BPI-1 and BPI-2. 
The shear rates are $\gd=1.95\e{-5}$ (red solid) 
and $\gd=3.91\e{-5}$ (green dashed). Compared to lower shear rates in BPI-1 
where the stress goes negative (Fig. \ref{bp1-fe-yield}) the stress is 
throughout positive, but exhibits very large fluctuations during one cycle, 
which are absent in regime BPI-2 at higher shear rates than shown here.}
\label{bp1-1_bp1-2}
\end{figure}

\begin{figure*}[htpb]
\includegraphics[width=0.245\textwidth]{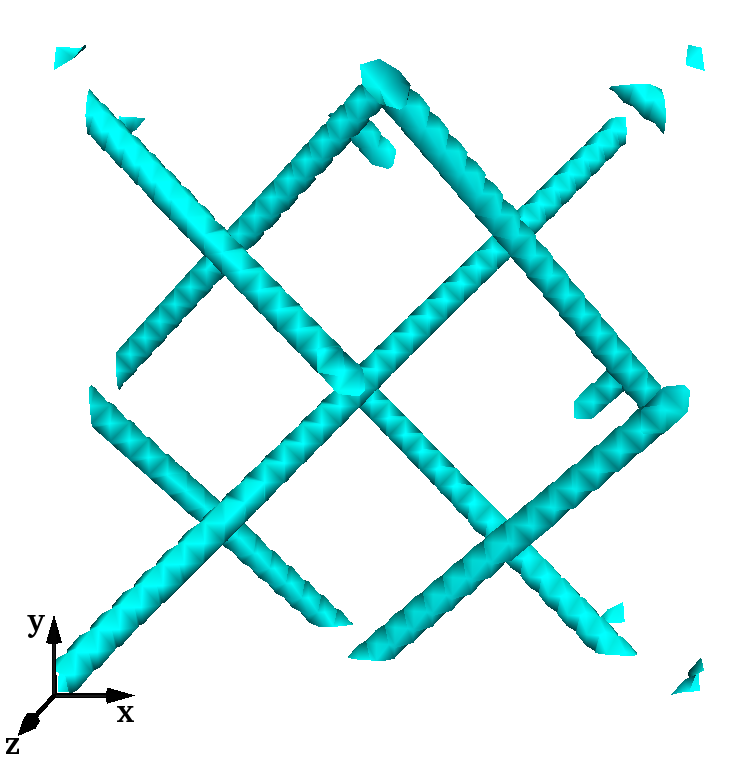}
\includegraphics[width=0.245\textwidth]{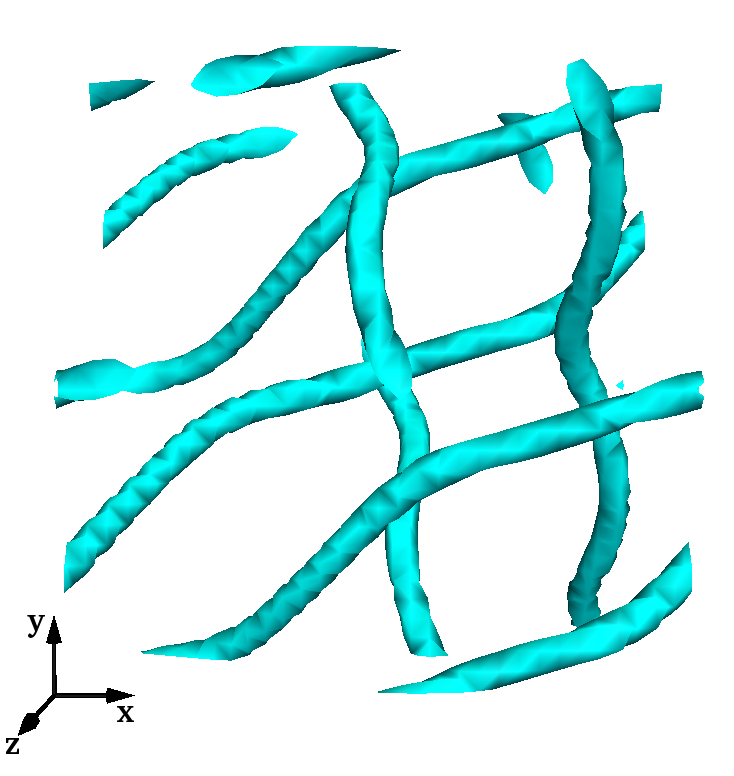}
\includegraphics[width=0.245\textwidth]{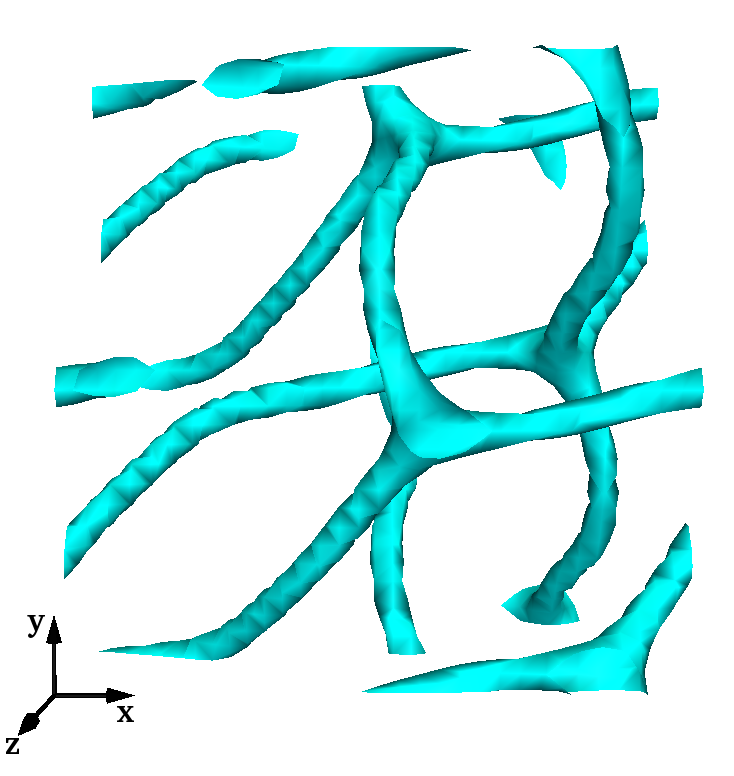}
\includegraphics[width=0.245\textwidth]{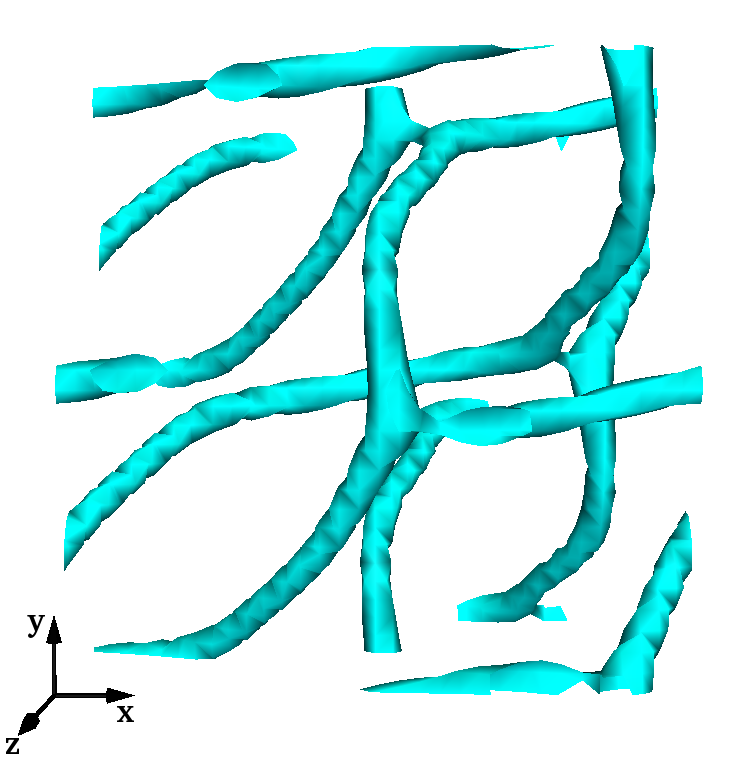}\\
\includegraphics[width=0.245\textwidth]{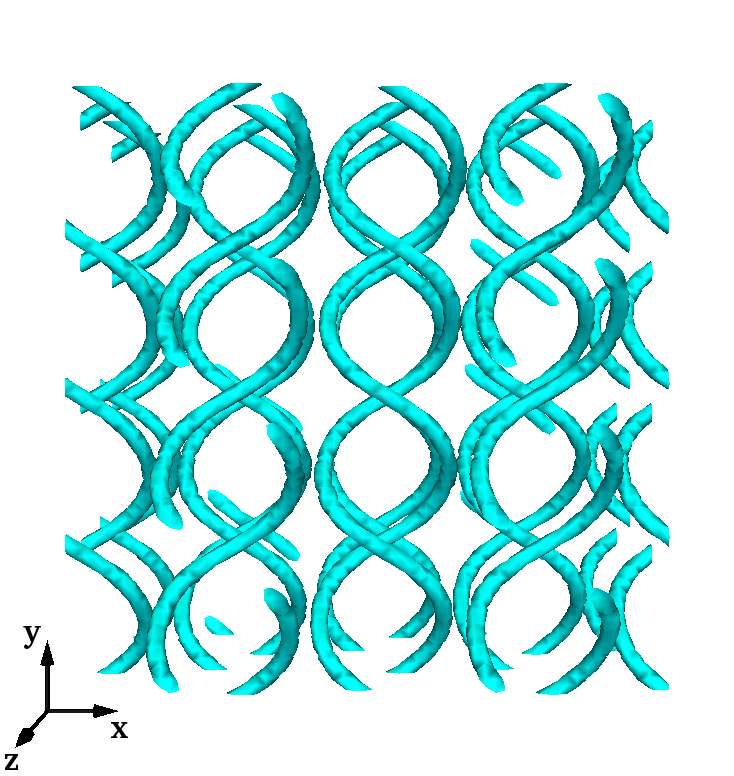}
\includegraphics[width=0.245\textwidth]{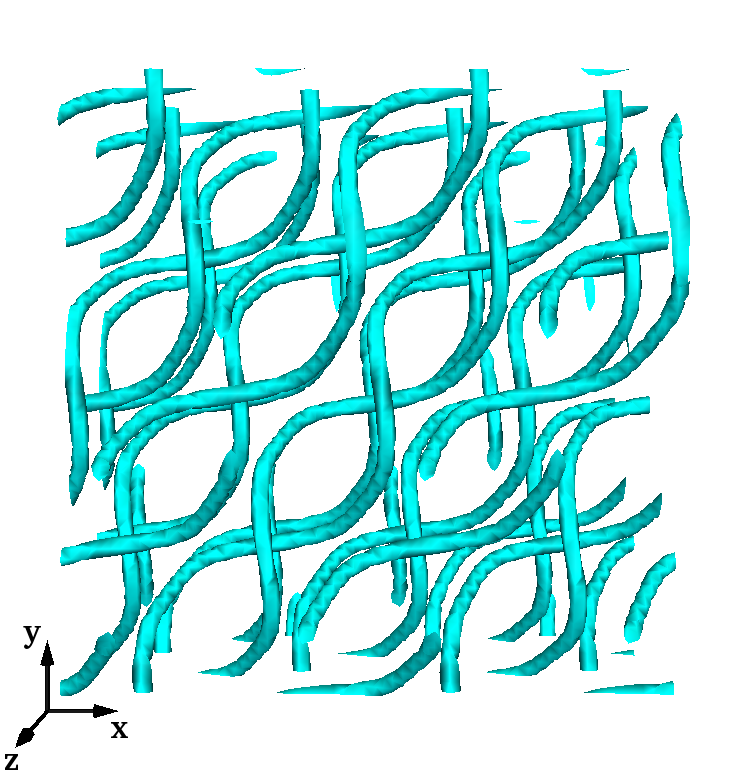}
\includegraphics[width=0.245\textwidth]{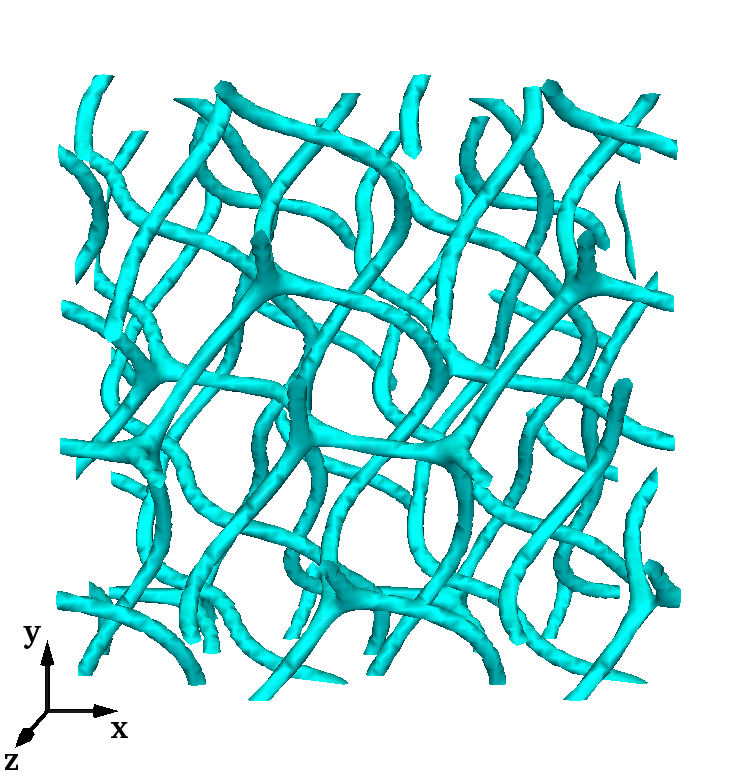}
\includegraphics[width=0.245\textwidth]{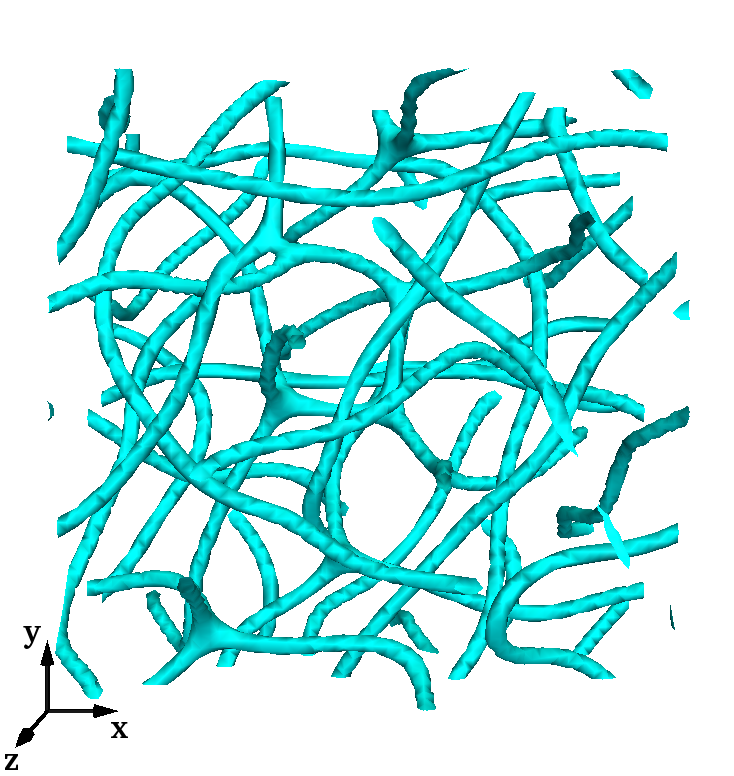}
\caption{Snapshots of BPI disclination network at $\dot{\gamma}=4.88\e{-6}$: 
Depicted is the transition from the quiescent state to flow-induced, intertwined 
helices that undergo a recurring structural transformation. The top row
shows a section of one unit cell for early times 
$t=1\e{4}, 1.8\e{5}, 2.0\e{5}$ and $2.1\e{5}$. The bottom 
pictures on the far left, centre left and centre right 
show the situation at later time steps $t=1.65, 1.8$ and $2.0\e{6}$,
which the system passes for several cycles before it ends up in
an amorphous state (bottom row far right, at time step $t=2.4\e{6}$).}
\label{bp1-1-disc}
\end{figure*}

To gain further insight into the flow-induced deformation, and eventual
breakdown, of BPI, it is instructive to follow the dynamics of 
the network under a slow flow -- this is depicted in Fig. \ref{bp1-1-disc}.
Three different stages can be distinguished. 
Just after the onset of the shear flow, the disclination lines 
in BPI get more and more squeezed together, and this is incompatible with
the defect topology in this phase.
Consequently, the network adopts a flow-induced conformation which consists 
of intertwined helices that stretch during the shear transformation.
This helical conformation that emerges already at strains $\gamma\simeq1$ 
is shown in Fig. \ref{bp1-1-disc} (top row).
At this point the original BPI has already been significantly deformed.
Shortly afterwards regular oscillations set up temporarily,
where the disclinations form double helices which tilt and realign under the shear.

This mode of flow proves unstable, as after a few cycles 
distortions appear which lead to further destabilisation -- presumably in
view of the large stress fluctuations discussed above. 
Finally, the system loses order and 
transforms into an amorphous network state with almost constant
stress in the region of $\Pi_{xy}\simeq 1-2\e{-5}$ LBU.
If the shear is stopped, this flow-induced amorphous state
remains arrested and metastable, and cannot
find its way back to the original BPI structure.

\begin{figure*}[htpb]
\includegraphics[width=0.32\textwidth]{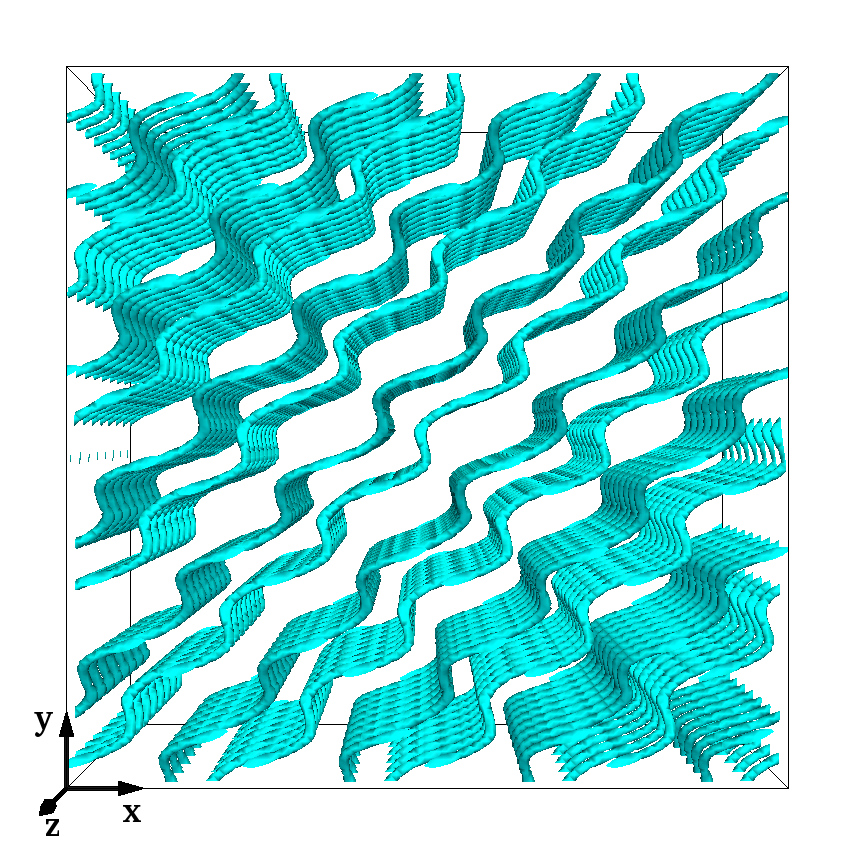}
\includegraphics[width=0.32\textwidth]{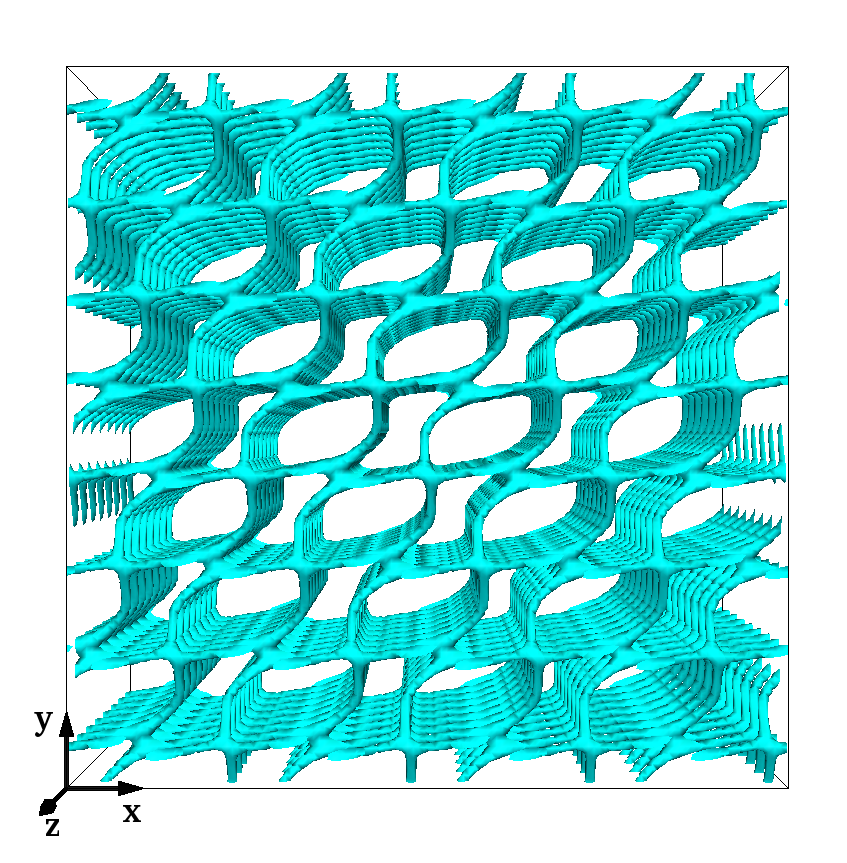}
\includegraphics[width=0.32\textwidth]{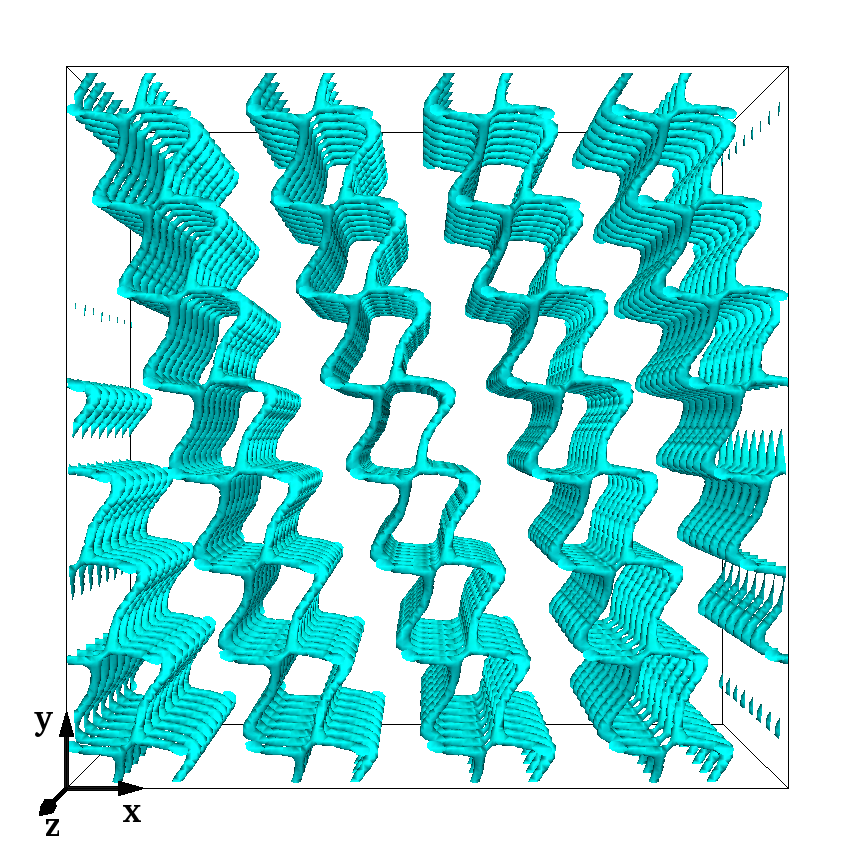}\\
\includegraphics[width=0.32\textwidth]{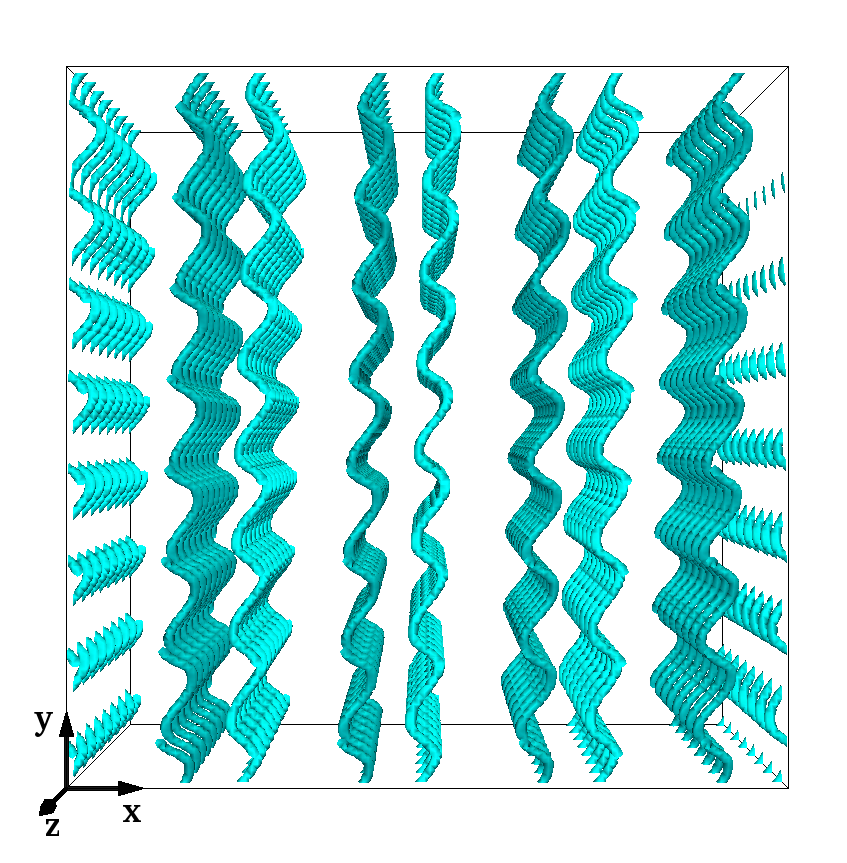}
\includegraphics[width=0.32\textwidth]{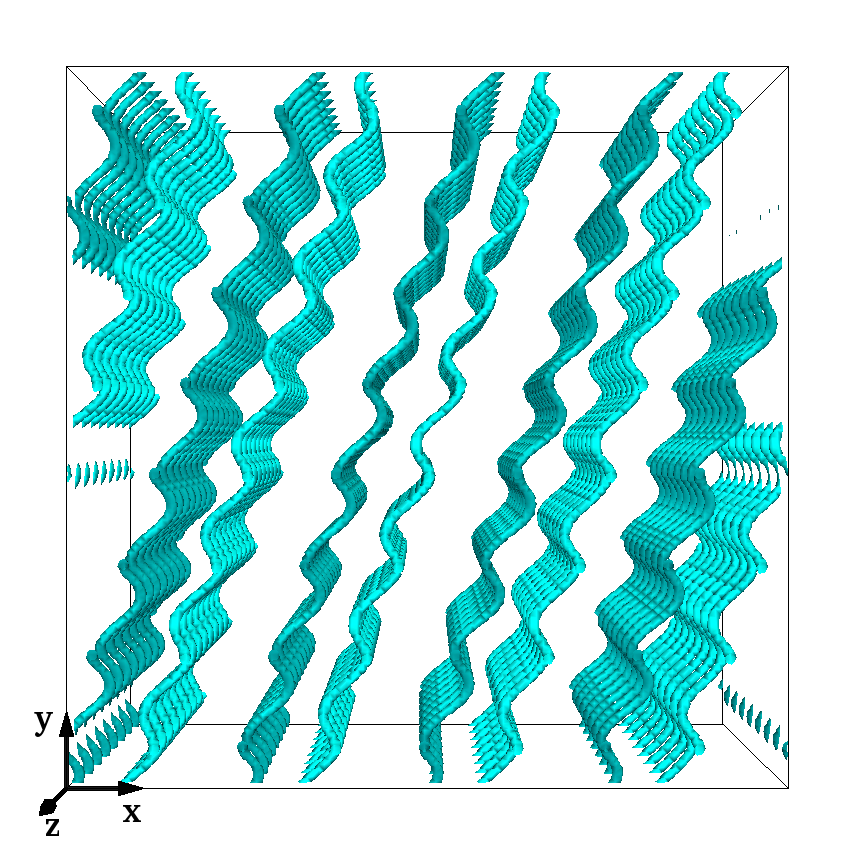}
\includegraphics[width=0.32\textwidth]{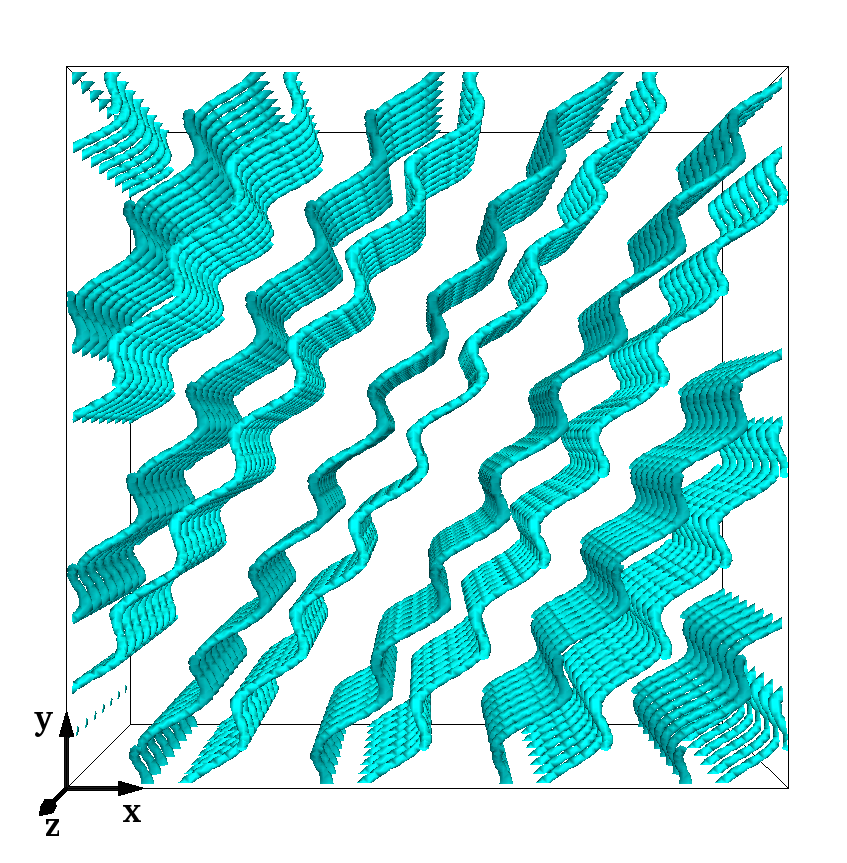}
\caption{Disclination network of BPI in regime BPI-2 (intermediate flow rates): 
The sequence shows a typical cycle of shear-induced transformations in the 
steady state at $\gd=1.56\e{-4}$ and time steps 
$t=3.64, 3.66,3.68,3.70,3.72,3.74\e{5}$ in flow-gradient plane. 
During every cycle the network also is displaced 
along the vorticity direction just as BPII
in regime BPII-1, but at a different rate.
}
\label{bp1-2-disc}
\end{figure*}

\subsubsection{Regime BPI-2: intermediate shear rates }

Adjacent to regime BPI-1 ($\gd\lesssim1.95\e{-5}$; $ {\it Er}\lesssim 0.33$) but 
at slightly larger shear rates ($3.91\e{-5}\lesssim\gd\lesssim 2.34\e{-4}$; $0.67\lesssim{\it Er}\lesssim2$)
lies another region where the network flows with periodically 
recurring conformations (Fig. \ref{bp1-2-disc}). 
We refer to this region as BPI-2. The transition between BPI-1 and BPI-2 
is clear when looking at Fig. \ref{bp1-1_bp1-2} 
which shows the shear stress versus total strain. 
A qualitative difference between these two is the absence in BPI-2 of the 
large stress fluctuations that occur during the early cycles in regime BPI-1.
(Recall that for even lower $\gd$ the thermodynamic shear stress becomes 
temporarily negative (Fig. \ref{bp1-fe-yield}), which 
caused destabilisation of the periodic network and led to the 
amorphous configuration in the steady state of BPI-1.)
Hence, an explanation
for the existence of the regular BPI-2 oscillations 
could be that compared to the large oscillations found transiently in
the BPI-1 regime these oscillations are now fast enough to bypass or suppress 
large stress fluctuations, which leads to
different topological reconnections and order structure
and eventually to a stabilisation of the flow. 

\begin{figure*}[htpb]
\includegraphics[width=0.32\textwidth]{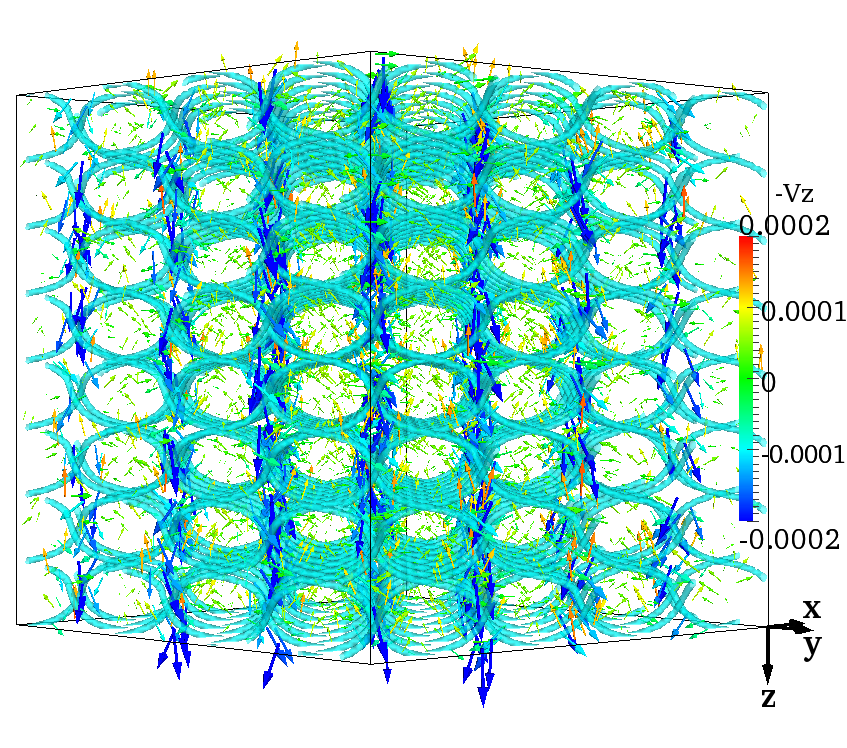}
\includegraphics[width=0.32\textwidth]{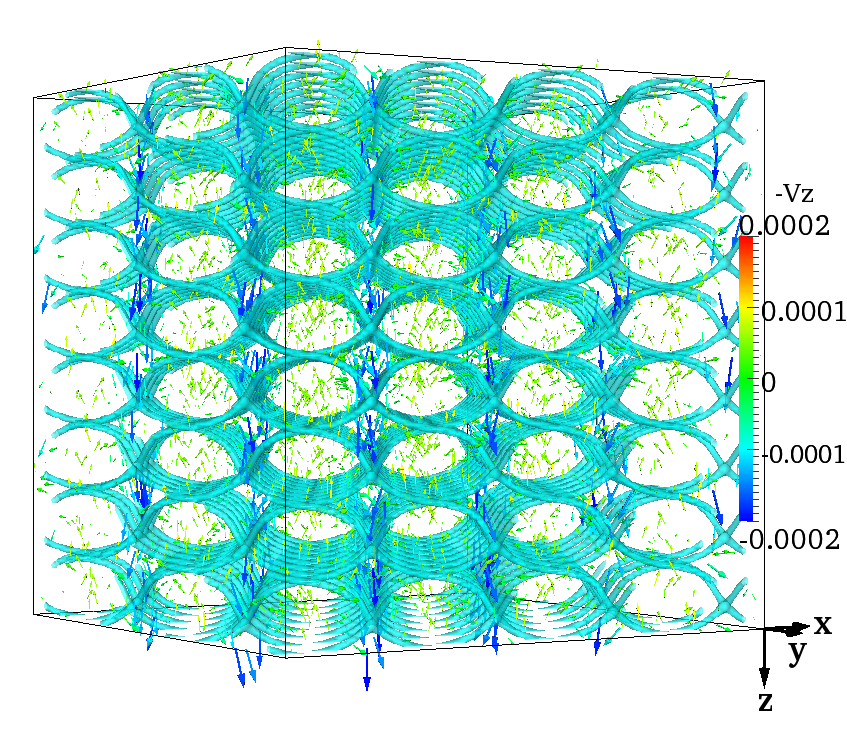}
\includegraphics[width=0.32\textwidth]{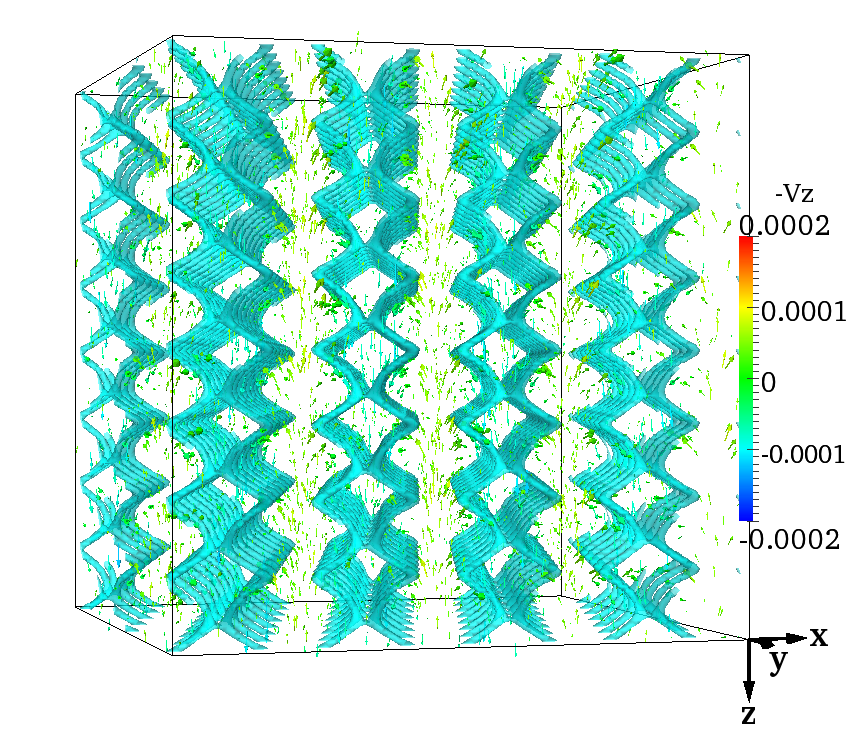}
\includegraphics[width=0.32\textwidth]{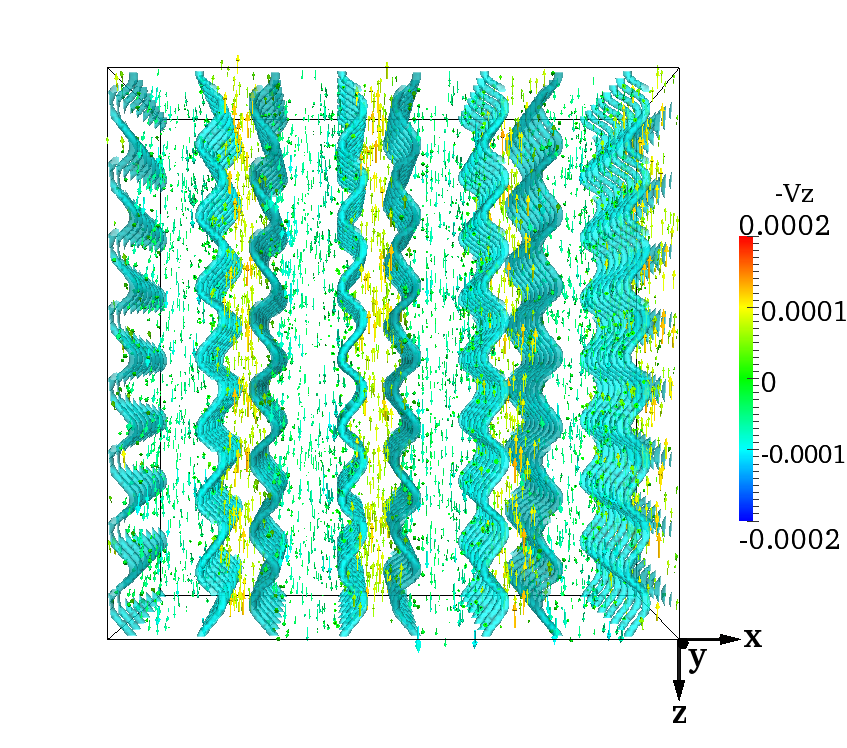}
\includegraphics[width=0.32\textwidth]{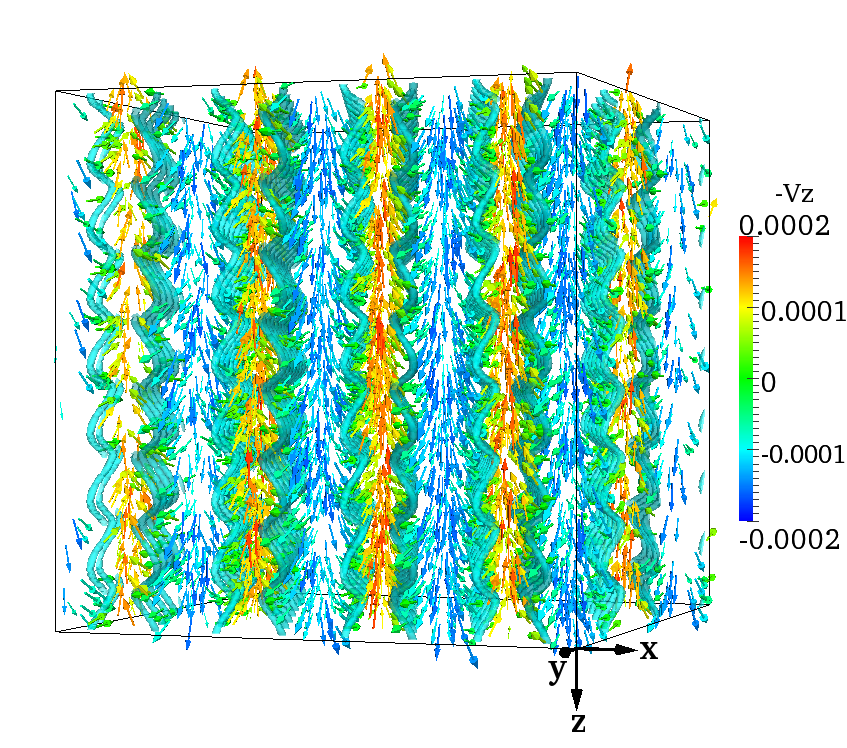}
\includegraphics[width=0.32\textwidth]{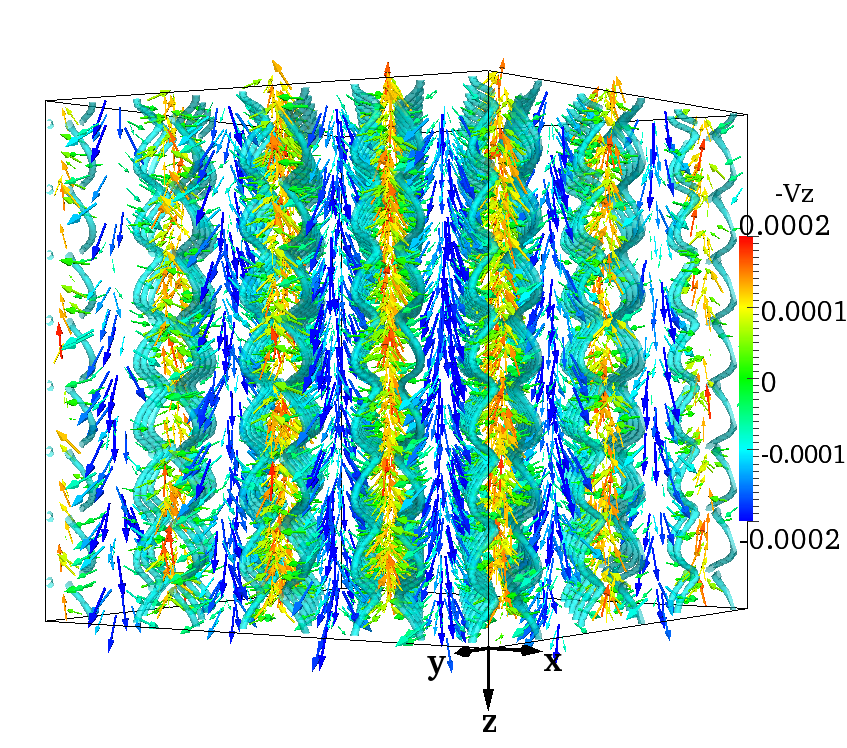}
\caption{Velocity patterns in BPI for positive helicity: 
The pictures show a snapshot of the periodically recurring patterns in the 
secondary velocity components $(v_y,v_z)$ at $\gd=1.56\e{-4}$ and time steps 
$t=3.64, 3.66,3.68,3.70,3.72,3.74\e{5}$, i.e. for the same shear rate
and time steps as in Fig.\ref{bp1-2-disc}. The colour code gives the magnitude and sign 
of the $z$-component. For negative helicity the sense of motion of the 
network and the secondary velocity components is inverted just as in Fig. \ref{bp2-1-velo}. 
Note that the viewing direction has been changed along an orbit in $xy$-plane
 to give a clearer view on the velocity pattern.
}
\label{bp1-2-velo}
\end{figure*}

Fig. \ref{bp1-2-disc} shows snapshots of the periodically recurring 
BPI in steady shear flow. Contrary to BPII-1 at these shear rates, 
BPI-2 does not resemble so much the equilibrium configuration
undergoing a homogeneous topology-conserving 
transformation. 
It features intricate, flow-induced 
conformations consisting of well-separated undulating 
disclination lines similar to those intertwined helices 
in Fig.~\ref{bp1-1-disc} at early times.

\begin{figure*}[htpb]
\includegraphics[width=0.32\textwidth]{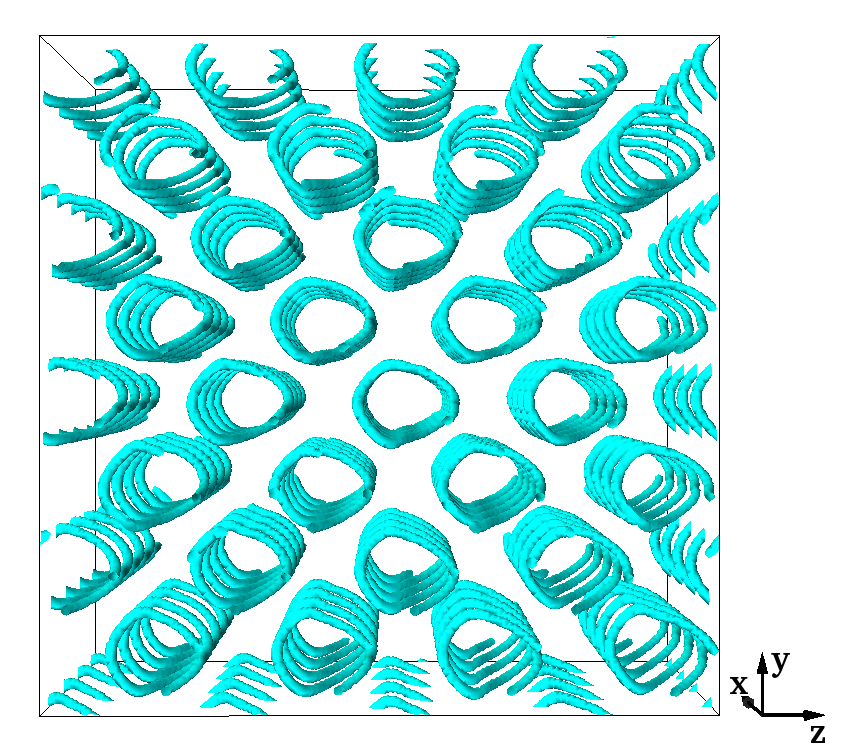}
\includegraphics[width=0.32\textwidth]{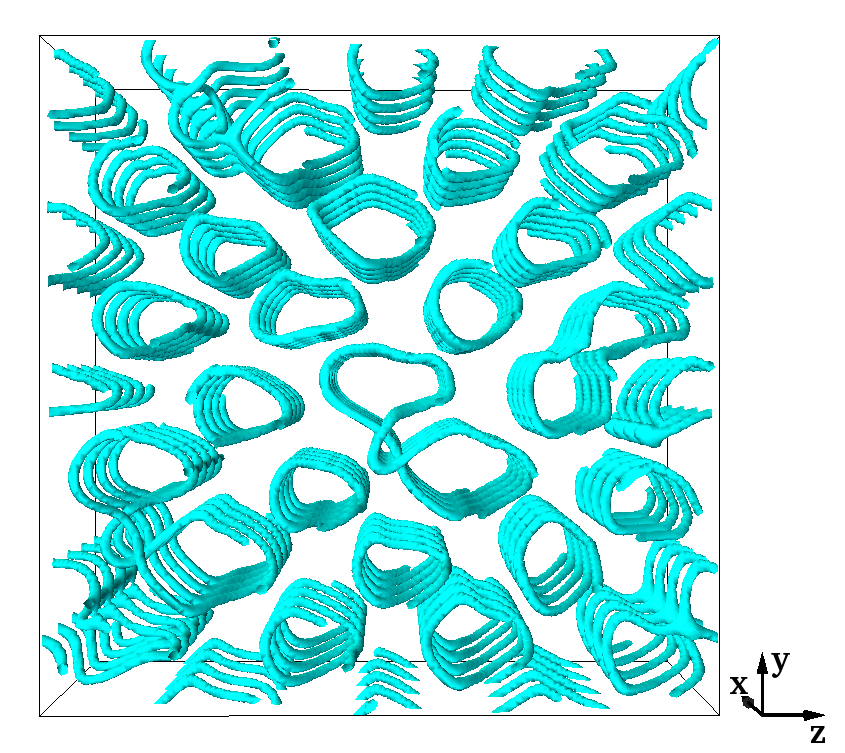}
\includegraphics[width=0.32\textwidth]{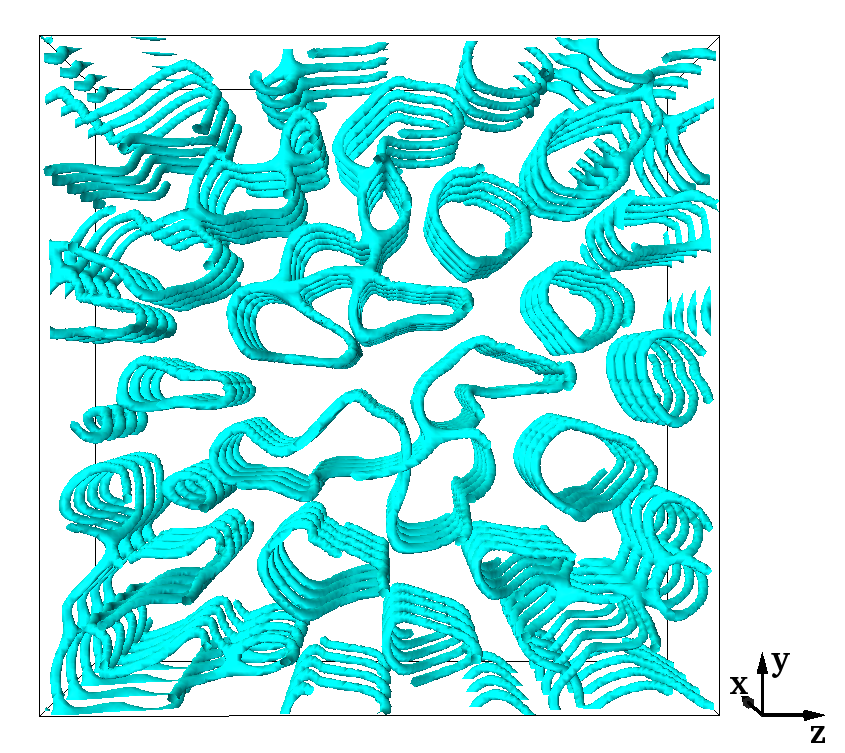}\\
\includegraphics[width=0.32\textwidth]{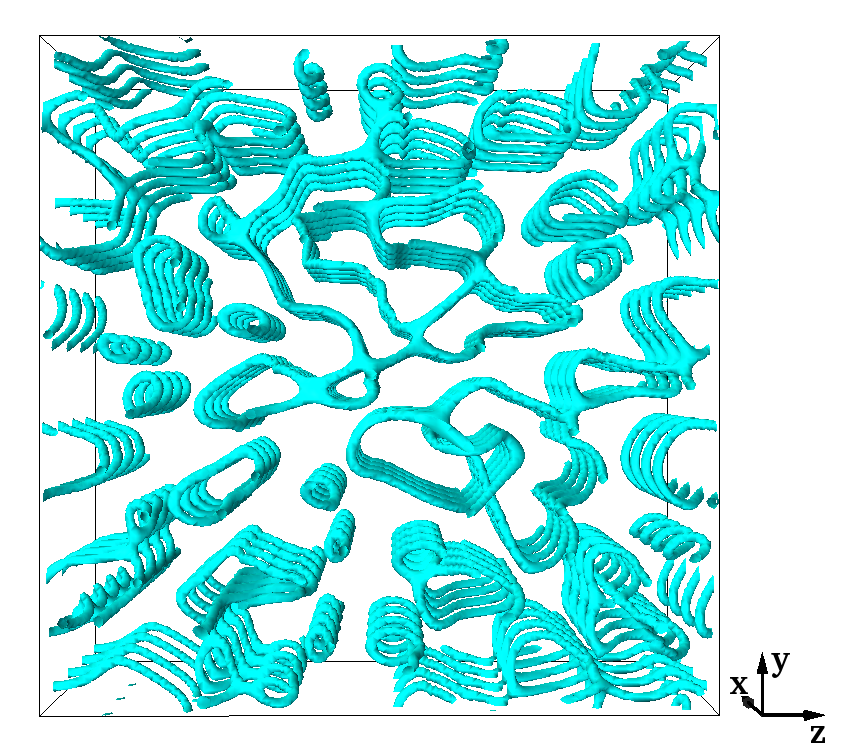}
\includegraphics[width=0.32\textwidth]{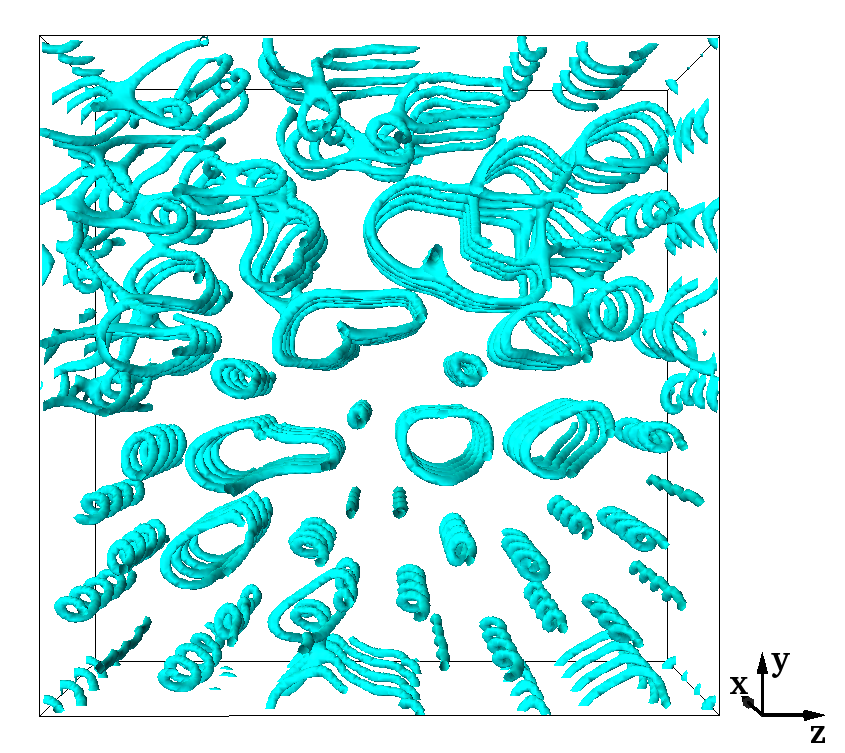}
\includegraphics[width=0.32\textwidth]{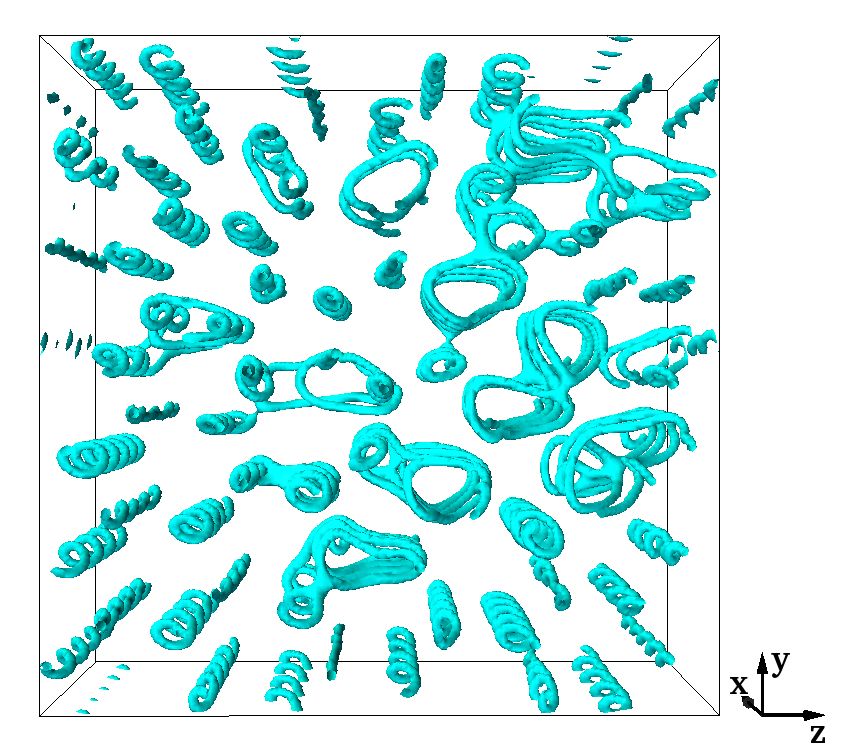}
\caption{
Disclination network of BPI in regime BPI-2 (high flow rates):
The sequence shows the break-up of the initially ordered state into 
an amorphous network at $\gd=3.125\e{-4}$ and time steps
$t=6.0, 6.75, 7.0, 7.25 \e{5}, 1.0,1.2\e{6}$ in vorticity-gradient plane.
During every cycle the network also is displaced
along the vorticity direction.
}
\label{bp1-3-disc}
\end{figure*}

If the shearing is switched off in
regime BPI-2 at some point during an oscillation, 
the flow-induced configuration cannot revert to a quiescent BPI.
Instead the network is trapped in a metastable state with 
helicoidal intertwined disclinations similar to those in the bottom left 
picture of Fig.~\ref{bp1-1-disc} or those in the top left picture of Fig.~\ref{bp1-3-disc}.
However, the orientation of the helical axis in now always 
along the vorticity ($z-$) direction.

Fig. \ref{bp1-2-velo} depicts a snapshot of the secondary velocity 
components $v_y$ and $v_z$ of BPI at the same time steps as in Fig.~\ref{bp1-2-disc}. 
The emerging pattern is similar to that of BPII shown in
Fig. \ref{bp2-1-velo}. 
There is also separation into fluctuating bands 
which are oriented along the vorticity direction.
The secondary velocity component $v_z$ changes sign upon
changing from positive to negative helicity and vice versa.
The recurrence period is different from the one we observe in BPII.
If we refer again to $\tau_F$ as the time interval between two consecutive 
reconnecting events of the network in flow direction, then $\tau_F=2/\gd$. 
A full reconnection in vorticity direction takes place after the time $\tau_V=4 \tau_F$.
We believe the reason for these differences is also directly linked to the different
topology of the defect network. The details of this mechanism will be
addressed in future work.

\subsubsection{Regime BPI-3: high shear rates }

At higher flow rates, $3.125\e{-4}\lesssim\gd\lesssim4.688\e{-4}$ 
($2.67\lesssim {\it Er}\lesssim 4$) we observe a break-up of the 
periodically recurring disclination network in regime BPI-2 and the formation
of another amorphous network. A sequence of snapshots of this is shown in Fig. \ref{bp1-3-disc}. 
At first the disclination lines take the form of a regular, staggered array of helices 
with the helical axis lying along the flow direction.
However, this mode of flow proves equally unstable as the regular formations dissolve into 
an irregular state with helicoidal disclination lines of varying 
appearance. The morphology is different from the amorphous network of
regime BPI-1 (cf. Fig.~\ref{bp1-1-disc}) where the disclination lines are 
well separated but do not have a clear orientation with respect to the flow. 

Interestingly, the critical Ericksen number for this observed break-up is in 
the range of $Er\simeq 2-4$. This is a regime where viscous forces have a 
similar magnitude to elastic forces and where the nonlinear coupling between
order and flow can cause small deviations to grow quickly over time. 
This suggests that the BPI-3 regime constitutes an instance 
of "rheochaos"~\cite{rheochaos,Cates:2002} as mentioned previously.

\subsection{Transition to cholesteric helix and flow-aligned nematic state}\label{gj-fan}

\begin{figure}[htpb]
\includegraphics[width=0.475\textwidth]{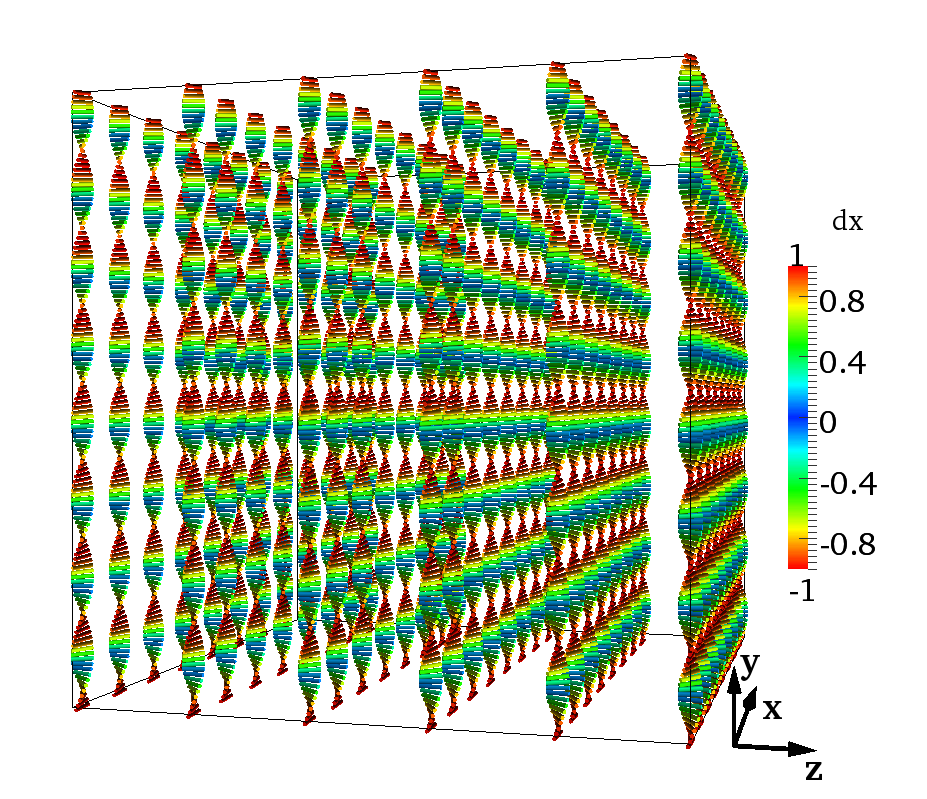}
\caption{Flow induced cholesteric state or Grandjean texture in regime BPI-4 and BPII-2: 
The orientation of the helix is along the gradient direction. 
The director field is quasi-static during the flow, which occurs 
in ``nematic layers''. Colour coding indicates the 
$x$-component of the director field (see colour bar on the right).
For clarity only selected sites are shown along the $y$-direction.
The structure is translationally invariant along the other two directions.}
\label{grandjean}
\end{figure}

At shear rates beyond the regimes BPI-3 and BPII-2, but below the transition to a 
flow-aligned nematic state at still higher shear rates, we found another regime where 
both blue phases adopt the same configuration in steady shear flow, 
independent of their initial state.
The director field of this configuration, also known as
Grandjean texture, is shown in Fig. \ref{grandjean}.
It consists of a simple cholesteric helix with the helical axis oriented 
along the gradient direction ($y$). While the liquid crystal is flowing 
the director field is stationary and 
retains its relative orientation so that the depicted state is translational invariant 
in flow-vorticity plane and quasi-one-dimensional. 
There is a small Leslie-type angle with respect to 
the flow-vorticity plane. 
Contrary to the other metastable flow states there is no tumbling motion 
of the director. This leads to lower shear stress and lower dissipation 
compared to other states like the travelling helical wave ~\cite{Rey:1996a, Rey:1996b}.

\begin{figure}[htpb]
\includegraphics[width=0.245\textwidth]{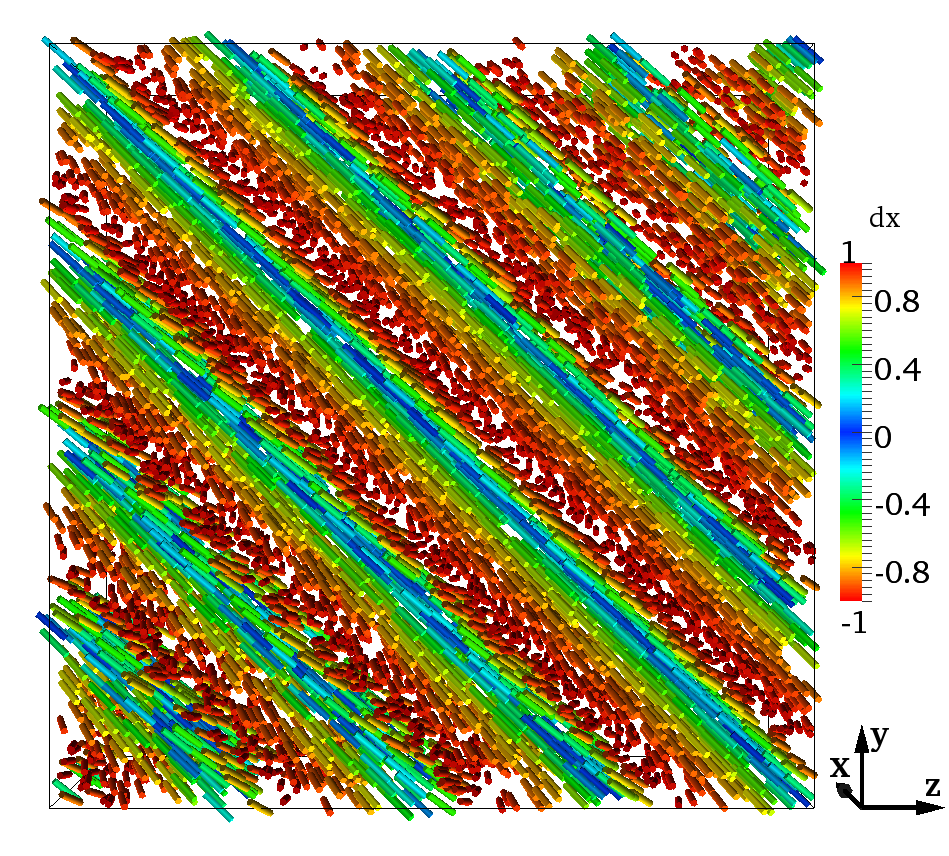}\nolinebreak
\includegraphics[width=0.245\textwidth]{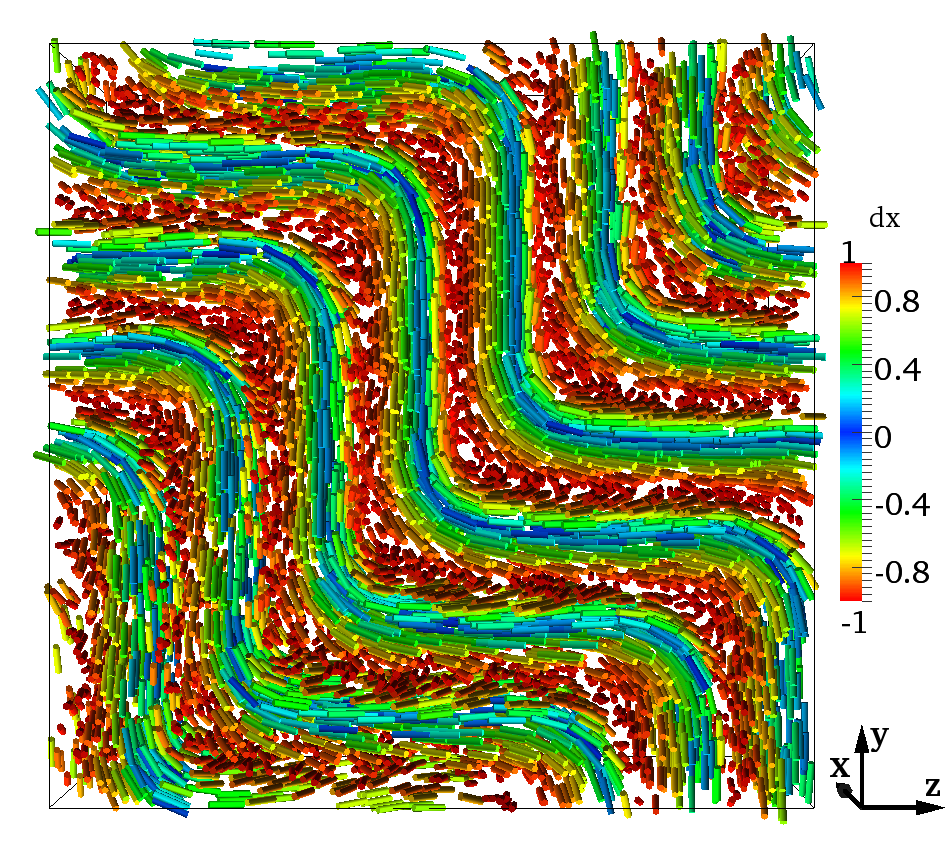}
\caption{
Director field of BPI-4 at high shear rate: The pictures show a frustrated 
Grandjean configuration that formed layers along a diagonal direction. 
The colour code gives the magnitude of the $x$-component in flow direction. 
}
\label{bp1-4-frust}
\end{figure}

The Grandjean configuration does not always emerge perfectly.
Fig. \ref{bp1-4-frust} shows a time sequence of a frustrated state 
that undergoes a buckling transition.
Initially the layers happen to be along a diagonal direction rather than
along the gradient direction. 
Because the diagonal state does not have the equilibrium layer spacing the conformation 
buckles which reduces its free energy.
Due to the periodic boundary conditions and their interlocking effect on the configuration 
this state cannot transform into a perfect Grandjean texture. However, the tilt angle of the
layers with respect to the flow-vorticity ($xz$) plane decreases significantly if the 
strain rate is increased.
This explains as well why the frustration is more frequent for low shear rates 
in regime BPI-4, as shown in Fig.~\ref{bp1-flowcurve}.

\section{Conclusions}

In summary, our work constitutes the first large scale simulation of bulk flow behaviour 
of cubic BPs in simple shear flow. 
  
We were able to characterise the rheology of cubic BPs, and identified 
three different flow regimes for BPII and five different ones for BPI.
These include some steady states that BPI and BPII have in common like the 
Grandjean texture and the flow-aligned nematic state at high shear rates.
Below an Ericksen number of $Er\simeq3$, BPII exhibits weak shear-thinning and 
obeys a power-law flow curve with exponent close to unity.  
The BPII disclination network breaks up and reconnects 
in the flow, which leads to a periodically recurring dependence 
of the shear stress. The flow-induced conformation looks generally
very similar to the quiescent network at equilibrium. 
While being homogeneously transformed due to the shear flow
the disclination network moves steadily in the vorticity direction
apparently by a permeation mechanism. The sense of
motion is directly linked to the helicity of the underlying cholesteric phase.

At larger Ericksen numbers, $4\lesssim Er\lesssim 10$, the flowing BPII network 
breaks up into a simple cholesteric helix with the helical axis 
along the gradient direction, also referred to as Grandjean texture. 
The travelling helical wave, predicted theoretically in Ref.~\cite{Rey:1996a,Rey:1996b}
under the assumption that the director is restricted to the flow-gradient plane,
turns out to be a metastable state of flow with a higher free energy.
It may also transforms into the Grandjean texture, the steady state with the lowest
shear stress in this range of shear rates.
Still larger flow rates break this residual
cholesteric order and leave a standard flow-aligned nematic state,
at the highest shear rate studied.

Interestingly, BPI shows a flow behaviour that is very different to that of BPII.
This is a direct consequence of the topological differences between their disclination 
networks. Below an Ericksen number of $Er\sim 0.3$, the BPI network cannot both flow and 
retain a regular appearance. It eventually reorganises into an amorphous network that
features yield-stress behaviour. The apparent reason is that shortly after the onset of the shearing 
very large stress fluctuations occur. At the lowest shear rates
the thermodynamic contribution to the shear stress becomes temporarily 
negative during each cycle. These fluctuations seem to destabilise the network and 
eventually trigger the transition into an amorphous state with a residual yield
stress. 

At slightly larger Ericksen numbers $0.4\lesssim Er \lesssim2$,
BPI kinematically bypasses the large stress fluctuations and 
flows with periodically recurring flow-induced conformations.
These conformations entail regularly arranged helical disclinations which 
rotate due to the shear. 
Despite their complex appearance, the flow-induced conformations are 
topologically connected to the quiescent BPI and after switching off the 
shear flow a defect-free blue phase reforms.

At even larger Ericksen numbers $2\lesssim Er \lesssim 4$ the behaviour of BPI is again different.
At first, it exhibits regular helicoidal disclination lines oriented along the flow direction.
After a short time these become irregular and an amorphous network emerges, which has 
a different appearance than that observed at the lowest flow rates.
We believe that at these flow rates, where neither viscous nor elastic forces clearly dominate, 
the coupling between order and flow makes any ordered conformation very sensitive towards 
irregular disturbances caused by small fluctuations.
It is tempting to
interpret the unsteady oscillations seen in this regime as an instance of 
deterministic rheological chaos~\cite{fielding, Cates:2002}.

Just as in the case of BPII, the BPI network breaks up at larger Ericksen numbers. 
First, at $5\lesssim\ Er\lesssim 16$, it forms a cholesteric helix along the flow gradient 
direction, the Grandjean texture. Finally, at $Er\gtrsim21$ (BPI) and $Er\gtrsim16$ (BPII) the configuration
is a flow-aligned nematic state.
Although experimental evidence to support
our results is currently not available, we hope this work will inspire such experiments, 
and believe it can shed some light on the flow properties of complex liquid-crystalline phases.

\section*{Acknowledgments}

This work was granted access to the HPC resource of CSC, Finland, made available 
within DECI by the PRACE-2IP, funded by the European Community's 7th Framework 
Program under grand agreement No. RI-283493. We also acknowledge support by 
EPSRC grant nos. EP/E045316 and EP/E030173 and the MAPPER EU-FP7 project 
(grant no. RI-261507). We thank Peter J. Collings and Tom C. Lubensky for stimulating discussions. 
MEC holds a Royal Society Research Professorship.

\appendix
\section*{Appendix}

Table \ref{tab1} lists the parameters chosen for our runs (values of
the reduced temperature and reduced chirality are given in the caption),
together with the minima, maxima, and standard deviations of the velocity
field in the three directions (discussed further in the main text). 
The first column also describes what flow regime each simulation
leads to in steady state.

In all our runs we used Lees-Edwards boundary conditions (LEBCs)~\cite{Wagner:2002}.
This is the sheared equivalent of periodic boundary conditions, for both the 
velocity and the order parameter field. 
On crossing a Lees-Edwards boundary a Galilean transformation is applied
with a velocity increment which is fixed in time (and limited in size by the
low Mach number constraint of LB). An appropriate transformation
of LB distributions which propagate across a boundary
must be applied at each time step, and appropriate
adjustment to the tensor of velocity gradients $W_{\alpha\beta}$ 
is required to compute cross-plane gradients of the velocity field 
used in the update to ${\mathbf Q}$. In both cases interpolation of the
relevant quantities is required to cope with the relative displacement
of neighbouring lattice sites separated by a sliding plane (the
displacement may be a fraction of a lattice unit at any given time
step) \cite{Henrich:2012a}. It is worth mentioning that apart from 
the Galilean transformation LEBCs do not impose any further constraint.
This means that order parameter field is free to follow its own kinetic pathway
at the Lees-Edwards plane. 
The use of multiple sliding planes equally spaced in a single
system allows the overall shear rate to be maintained indefinitely as the system
becomes larger in the velocity gradient direction (in contrast with the use of solid
walls to impose shear).
While extremely useful to simulate bulk flow, we
note that these boundary conditions impose macroscopic distortion
of the network. They do not allow for a free-standing network 
sustained stationary by permeation along the flow direction.
Such a state is however forbidden for large system sizes since the 
local permeation rate would have to increase indefinitely with 
sample thickness.
The boundary condition is equivalent to infinitely distant
walls along the $xy-$plane where the liquid crystal is anchored,
not unlike the situation found in a rheometer with no-slip
boundary conditions at the walls.
In practice, permeation flows are 
only restricted to very low flow rates, and are unstable for intermediate 
and fast flows~\cite{Dupuis:2005}, 
where the response of the network depends less on
the details of the boundary conditions used at the wall.
With free boundary conditions that do not impose a macroscopic distortion 
and for very low flow velocity, 
permeative flows might instead allow for
some slip of the BPI network, as in cholesterics sheared
along their axis~\cite{Marenduzzo:2006a, Marenduzzo:2006b}.
However, as mentioned above, a triply periodic structure cannot
be maintained stationary in a large system without infinite 
permeation rate.

\begin{table*}[htpb]
\begin{tabular}{|c||c|| c || c || c |c |c||c| c| c||c| c| c|}
\hline
& $\dot{\gamma}$ & ${\it Er}$ & $q_0$ & $\bar{v}_{x,min}$ & $\bar{v}_{x,max}$ & $\bar{v}_{x,std}$ & $\bar{v}_{y,min}$ & $\bar{v}_{y,max}$ & $\bar{v}_{y,std}$ & $\bar{v}_{z,min}$ & $\bar{v}_{z,max}$ & $\bar{v}_{z,std}$ \\
\hline
BPI & & & & $\times 10^5$\\
\hline
BPI-1 (AN) &0.24 &0.02 & 0.1388 &-17.7 &16.7 &3.7 &-3.6 &3.6 &3.5 &-3.9 &3.1 &2.8 \\
BPI-1 (AN) &0.49 &0.04 & 0.1388 &-33.4 &32.0 &4.3 &-3.0 &3.1 &3.2 &-3.7 &2.5 &3.1 \\
BPI-1 (AN) &0.98 &0.08 & 0.1388 &-64.4 &63.6 &4.7 &-3.6 &3.4 &4.8 &-5.3 &-4.2 &4.4 \\
BPI-1 (AN) &1.95 &0.17 & 0.1388 &-126.2 &127.0 &6.8 &-3.8 &4.8 &6.7 &-5.2 &5.1 &6.2 \\
\hline
BPI-2 (PRC) &3.91 &0.33 & 0.1388 &-248.6 &248.3 &2.8 &-5.4 &5.6 &5.1 &-3.8 &4.6 &3.9 \\
BPI-2 (PRC) &7.81 &0.67 & 0.1388 &-497.5 &496.4 &5.2 &-8.5 &8.0 &8.4 &-5.9 &7.8 &6.8 \\
BPI-2 (PRC) &15.63 &1.33 & 0.1388 &-995.0 &993.0 &8.7 &-13.1 &13.8 &13.2 &\bf{-13.0} &\bf{15.6} &\bf{11.2} \\
BPI-2 (PRC) &15.63 &1.33 & -0.1388 &-995.0 &993.0 &8.7 &-13.1 &13.8 &13.2 &\bf{-15.6} &\bf{13.0} &\bf{11.2} \\
BPI-2 (PRC) &23.44 &2.00 & 0.1388 &-1492.2 &1485.5 &9.6 &-4.8 &3.7 &12.3 &-5.7 &4.6 &14.8 \\
\hline
BPI-3 (AN) &31.25 &2.67 & 0.1388 &-1990.6 &1983.2 &17.6 &-4.5 &6.3 &13.3 &-8.3 &10.0 &15.9 \\
BPI-3 (AN) &46.87 &4.00 & 0.1388 &-2988.2 &2985.5 &42.4 &-10.5 &12.3 &28.4 &-15.5 &15.2 &26.8 \\
\hline
BPI-4 (GJ) &54.69 &4.67 & 0.1388 &-3475.1 &3473.4 &- &- &- &- &-8.9 &8.8 &- \\
BPI-4 (FGJ) &62.50 &5.33 & 0.1388 &-4101.7 &4122.7 &- &- &- &- &-29.6 &20.0 &- \\
BPI-4 (FGJ) &78.13 &6.01 & 0.1388 &-5140.0 &5092.3 &- &- &- &- &-35.8 &24.6 &- \\
BPI-4 (GJ) &93.75 &8.00 & 0.1388 &-5955.4 &5946.1 &-  &- &- &- &-15.5 &15.1 &- \\
BPI-4 (GJ) &125.0 &10.67 & 0.1388 &-7935.0 &7933.7 &- &- &- &- &-21.3 &20.1 &- \\
BPI-4 (GJ) &187.5 &16.00 & 0.1388 &-11901.7 &11901.7 &- &- &- &- &-32.8 &29.1 &- \\
\hline
BPI-5 (FN) & 250.0 &21.33 & 0.1388 &-15875.0 &15875.0 &- &- &- &- &-0.2 &-0.1 &- \\
BPI-5 (FN) & 375.0 &32.00 & 0.1388 &-23812.5 &23812.5 &- &- &- &- &-0.3 &0.1 &- \\
\hline

BPII  & & & & $\times 10^5$ \\
\hline
BPII-1 (PRC) &0.24 &0.02 & 0.1963 &-7.8 &7.7 &0.1 &-0.2 &0.2 &0.1 &-0.3 &0.3 &0.2 \\
BPII-1 (PRC) &0.49 &0.04 & 0.1963 &-15.5 &15.4 &0.2 &-0.3 &0.4 &0.2 &-0.5 &0.4 &0.2 \\
BPII-1 (PRC) &0.98 &0.08 & 0.1963 &-31.0 &30.8 &0.3 &-0.6 &0.6 &0.4 &-0.9 &0.8 &0.4 \\
BPII-1 (PRC) &1.95 &0.17 & 0.1963 &-62.0 &61.6 &0.6 &-1.2 &1.2 &0.8 &-1.8 &1.6 &0.8 \\
BPII-1 (PRC) &3.91 &0.33 & 0.1963 &-123.9 &123.3 &1.2 &-2.1 &2.1 &1.6 &-3.4 &3.0 &1.5 \\
BPII-1 (PRC) &7.81 &0.67 & 0.1963 &-247.5 &246.8 &2.0 &-3.8 &3.8 &3.0 &-6.5 &5.5 &2.6 \\
BPII-1 (PRC) &15.63 &1.33 & 0.1963 &-494.3 &493.7 &2.5 &-5.5 &-5.5 &4.0 &\bf{-10.9} &\bf{8.7} &\bf{3.9} \\
BPII-1 (PRC) &15.63 &1.33 & -0.1963 &-494.3 &493.7 &2.5 &-5.5 &5.5 &4.0 &\bf{-8.7} &\bf{10.9} &\bf{3.9} \\
BPII-1 (PRC) &31.25 &2.67 & 0.1963 &-987.5 &987.1 &3.5 &-6.9 &6.9 &5.1 &-17.7 &13.1 &5.0 \\
BPII-1 (PRC) &39.06 &3.33 & 0.1963 &1233.9 &1233.6 &3.7 &-6.9 &6.9 &5.1 &-20.0 &14.5 &5.2 \\
\hline
BPII-2 (GJ) &46.87 &4.00 & 0.1963 &-1479.8 &1477.0 &- &- &- &- &-10.0 &9.6  &- \\
BPII-2 (GJ) &54.69 &4.67 & 0.1963 &-1720.5 &1720.8 &- &- &- &- &-11.5 &11.2 &- \\
BPII-2 (GJ) &62.5 &5.33 & 0.1963  &-1966.7 &1966.2 &- &- &- &- &-13.3 &12.7 &- \\
BPII-2 (GJ) &78.13 &6.01 & 0.1963 &-2458.3 &2457.8 &- &- &- &- &-16.9 &15.7 &- \\
BPII-2 (GJ) &93.75 &8.00 & 0.1963 &-2949.7 &2949.7 &- &- &- &- &-20.3 &18.4 &- \\
BPII-2 (GJ) &125.0 &10.67 &0.1963 &-3932.8 &3932.8 &- &- &- &- &-26.6 &22.4 &- \\
\hline
BPII-3 (FN) &187.5 &16.00 & 0.1963 &-5906.2  &5906.2 &- &-  &-  &-  &-0.6 &0.4 &- \\
BPII-3 (FN) &250.0 &21.33 & 0.1963 &-7874.9  &7874.9 &- &-  &-  &-  &-0.5 &0.3 &- \\
\hline
\end{tabular}
\caption{
Minima, maxima and standard deviation of time-averaged velocity components 
for BPI ($\tau=-0.5, \kappa=1.0$) and BPII ($\tau=-0.5, \kappa=2.0$) in steady 
shear flow. All velocity values are given in $10^{-5}$ lattice units. 
The last six columns refer to the secondary flow.
The transient dynamics
prior to regular oscillations or network break-up has not been included into the averages. 
The regimes comprise the formation of an amorphous network (AN, BPI-1), 
periodically recurring conformations with oscillatory stress response (PRC, BPI-2 and BPII-1), 
amorphous networks at Ericksen numbers ${Er\simeq O(1)}$ (AN, BPI-3), 
a Grandjean texture that may be frustrated (GJ \& FGJ, BPII-2 and BPI-4),
and a flow-aligned nematic state (FN, BPI-5 and BPII-3).
Bold figures indicate mean minima and maxima in $v_z$ and the standard deviation of the runs with 
inverted helicity. 
}
\label{tab1}
\end{table*}

\footnotesize{
\bibliography{bprheo_condmat} %your .bib file
\bibliographystyle{rsc} %the RSC's .bst file
}

\end{document}